\documentclass[aps,preprint,floatfix,showpacs,superscriptaddress]{revtex4}

\usepackage{graphicx}
\usepackage{amssymb}
\usepackage{bm}

\newcommand{\be}{\begin{equation}}
\newcommand{\ee}{\end{equation}}
\newcommand{\bel}[1]{\be\label{#1}}
\newcommand{\re}[1]{Eq.~(\ref{#1})}
\newcommand{\ds}{\displaystyle}
\newcommand{\ov}[1]{\overline{#1}}
\newcommand{\hsp}{\hspace*{1pt}}

\begin{document}

\title{1+1 Dimensional Hydrodynamics for High--energy\\
Heavy--ion Collisions}

\author{L.M. Satarov}

\affiliation{Frankfurt Institute for Advanced Studies,
J.W.~Goethe University, Max--von--Laue--Str.~1,~D--60438 Frankfurt
am Main, Germany}

\affiliation{The Kurchatov Institute, Russian Research Center,
123182 Moscow, Russia}

\author{I.N.~Mishustin}

\affiliation{Frankfurt Institute for Advanced Studies,
J.W.~Goethe University, Max--von--Laue--Str.~1,~D--60438 Frankfurt
am Main, Germany}

\affiliation{The Kurchatov Institute, Russian Research Center,
123182 Moscow, Russia}

\author{A.V.~Merdeev}

\affiliation{The Kurchatov Institute, Russian Research Center,
123182 Moscow, Russia}

\author{H.~St\"ocker}

\affiliation{Frankfurt Institute for Advanced Studies,
J.W.~Goethe University, Max--von--Laue--Str.~1,~D--60438 Frankfurt
am Main, Germany}

\begin{abstract}
A 1+1 dimensional hydrodynamical model in the light--cone coordinates
is used to describe central heavy--ion collisions at ultrarelativistic
bombarding energies. Deviations from Bjorken's scaling are taken into
account by choosing finite--size profiles for the initial energy
density. The sensitivity of fluid dynamical evolution to the equation
of state and the parameters of initial state is investigated.
Experimental constraints on the total energy of produced particles are
used to reduce the number of model parameters.  Spectra of secondary
particles are calculated assuming that the transition from the
hydrodynamical stage to the collisionless expansion of matter occurs
at a~certain freeze--out temperature. An important role of resonances in
the formation of observed hadronic spectra is demonstrated. The
calculated rapidity distributions of pions, kaons and antiprotons in
central Au+Au collisions at $\sqrt{s_{\scriptscriptstyle NN}}=200$\,GeV
are compared with
experimental data of the BRAHMS Collaboration. Parameters of the initial
state are reconstructed for different choices of the equation of state.
The best fit of these data is obtained for a soft equation of state and
Gaussian--like initial profiles of the energy density, intermediate
between the Landau and Bjorken limits.
\end{abstract}


\maketitle


\section{Introduction}

High--energy heavy--ion collisions provide a unique tool for studying
properties of hot and dense strongly--interacting matter in the
laboratory. The theoretical description of such collisions is often
done within the framework of a hydrodynamic approach.  This approach
opens the possibility to study the sensitivity of collision dynamics
and secondary particle distributions to the equation of state (EOS) of
the produced matter. The two most famous realizations of this approach,
which differ by the initial conditions, have been proposed by
Landau~\cite{Lan53} (full stopping) and Bjorken~\cite{Bjo83} (partial
transparency). In recent decades many versions of the hydrodynamic
model were developed, ranging from simplified
1+1~\cite{Mel58,Tar77,Mis83,Bla87,Esk98,Moh03} and 2+1 dimensional
models~\cite{Bla87,Kol99,Bas00,Per00,Kol01,Tea01} of the Landau or
Bjorken type to more sophisticated  3+1 dimensional models
\cite{Sto80,Ris95a,Non00,Hir02,Ham05,Non05}. One should also mention
the multi--fluid models
\cite{Ams78,Cla86,Bar87,Mis88,Kat93,Bra00,Ton03} which consider the
whole collision process including the nuclear interpenetration stage.
Recent theoretical investigations show that fluid--dynamical models
give a very good description of many observables at the SPS and RHIC
bombarding energies (see e.g. Ref.~\cite{Sto05}).

The 2+1 dimensional hydrodynamical models have been successfully
applied \mbox{\cite{Kol99,Bas00,Per00,Kol01,Tea01}} to describe the
$p_T$ distributions of mesons and their elliptic flow at midrapidity.
These models assume a boost--invariant expansion~\cite{Bjo83} of matter
in the longitudinal (beam) direction and, therefore, cannot explain
experimental data in a broad rapidity region, where strong deviations
from the scaling regime have been observed. More realistic 3+1
dimensional fluid--dynamical simulations have been already performed
for heavy--ion collisions at SPS and RHIC energies. But as a rule, the
authors of these models do not study the sensitivity of results to
the choice of initial and final (freeze--out) stages. On the other
hand, it is not clear at present, which initial conditions,
Landau--like~\cite{Lan53} or Bjorken--like~\cite{Bjo83}, are more
appropriate for ultrarelativistic collisions.

Our main goal in this paper is to see how well the fluid--dynamical
approach describes the RHIC data on $\pi, K, \overline{p}$
distributions over a broad rapidity interval, reported recently by the
BRAHMS Collaboration~\cite{Bea04,Bea05}. Within our approach we
explicitly impose a constraint on the total energy of the produced
particles which follows from these data.

For our study we apply a simplified version of the hydrodynamical
model, dealing only with the longitudinal dynamics of the fluid. This
approach has as its limiting cases the Landau and Bjorken models. We
investigate the sensitivity of the hadron rapidity spectra to the
fluid's equation of state, to the choice of initial state and
freeze--out conditions. Modification of these spectra due to the
feeding from resonance decays is also analyzed. Special attention is
paid to possible manifestations of the deconfinement phase transition.
In particular, we compare the dynamical evolution of the fluid with and
without the phase transition.

The paper is organized as follows: in Sect.~\ref{form} we formulate the
model and specify the equation of state, the initial and freeze-out
conditions. In this section we also explain how we calculate particle
spectra and take into account feeding from resonance decays. In
Sect.~\ref{sres} we present the numerical results. We first consider
the evolution of fluid--dynamical profiles for different equations of
states and initial conditions. Then these results are used to
calculate spectra of pions, kaons and antiprotons. In the end of this
section we analyze the initial conditions motivated by the Landau
and Bjorken models. Summary and outlook are given in Sect.~\ref{scon}.
Some details of numerical scheme and the procedure to calculate
feeding from resonance decays are given in Appendix.

A short version of this paper is published in Ref.~\cite{Sat06}.

\section{Formulation of the model}\label{form}

\subsection{Equations of ideal fluid dynamics\label{feqs}}

Below we study the evolution of highly excited, and possibly
deconfined, strongly--interacting matter produced in ultrarelativistic
heavy--ion collisions. It is assumed that after a certain
thermalization stage this evolution can be described by the
relativistic fluid dynamics. Conservation laws of the baryon charge and
4--momentum are expressed by the following differential
equations~\cite{Lan59}:
\begin{eqnarray}
\label{bcce}
&&\partial_\mu\left(n\hsp U^\mu\right)=0\,,\\
\label{emte}
&&\partial_\nu T^{\mu\nu}=0\,,
\end{eqnarray}
where $n,U^\mu$ and $T^{\mu\nu}$ are the rest--frame baryon
number density, the collective 4--velocity and the energy--momentum
tensor of the fluid. In the limit of small dissipation,
one can represent $T^{\mu\nu}$ in the standard form\footnote
{
Units with $\hbar=c=1$ are used throughout the paper.
}
\bel{enmt}
T^{\mu\nu}=(\epsilon+P)\hsp U^\mu U^\nu -P\hsp g^{\mu\nu}.
\ee
Here $\epsilon$ and $P$ are the rest--frame energy density and pressure.

We consider central collisions of equal nuclei disregarding the effects
of transverse collective expansion. In this case one can parametrize
$U^\mu$ in terms of the longitudinal flow rapidity~$Y$ as
$U^\mu=(\cosh{Y},\bm{0},\sinh{Y})^\mu$. It is convenient to make
transition from the usual space--time coordinates $t,z$ to
the hyperbolic (light--cone) variables~\cite{Bjo83}, namely,
the proper time~$\tau$ and the space--time rapidity $\eta$\,, defined as
\bel{lcva}
\tau=\sqrt{t^2-z^2}\,,\hspace*{5mm}\eta=\tanh^{-1}\left(\frac{z}{t}\right)=
\frac{1}{2}\ln{\frac{t+z}{t-z}}\,.
\ee
In these coordinates the equations (\ref{bcce})--(\ref{enmt}) take the
following form~\cite{Bel96}
\begin{eqnarray}
\left[\tau\frac{\partial}{\partial\tau}+
\tanh\hsp (Y-\eta)\hsp\frac{\partial}{\partial\eta}\right]\hsp n
+n\left[\tanh\hsp (Y-\eta)\hsp\tau\frac{\partial}{\partial\tau}+
\frac{\partial}{\partial\eta}\right] Y&=&0\,,
\label{bcce1}\\
\left[\tau\frac{\partial}{\partial\tau}+
\tanh\hsp (Y-\eta)\hsp\frac{\partial}{\partial\eta}\right]\hsp\epsilon
+(\epsilon +P)\left[\tanh\hsp (Y-\eta)\hsp\tau\frac{\partial}{\partial\tau}+
\frac{\partial}{\partial\eta}\right] Y&=&0\,,
\label{emte1}\\
(\epsilon +P)\left[\tau\frac{\partial}{\partial\tau}+\tanh\hsp
(Y-\eta)\hsp\frac{\partial}{\partial\eta}\right] Y
+\left[\tanh\hsp (Y-\eta)\hsp\tau\frac{\partial}{\partial\tau}+
\frac{\partial}{\partial\eta}\right] P&=&0\,.
\label{emte2}
\end{eqnarray}
To solve Eqs.~(\ref{bcce1})--(\ref{emte2}), one needs to specify the
EOS, \mbox{$P=P\hsp (n,\epsilon)$}, and the initial profiles
$n\hsp (\tau_0,\eta),\epsilon\hsp (\tau_0,\eta), Y(\tau_0,\eta)$
at a time $\tau=\tau_0$\, when the fluid may be considered as
thermodynamically equilibrated.

In this paper we consider only the baryon--free matter, i.e. assume
vanishing net baryon density $n$ and chemical potential $\mu$. In this
case Eq.~(\ref{bcce1}) is trivially satisfied and all thermodynamic
quantities, e.g. pressure, temperature $T$ and entropy density
$s=(\epsilon+P-n\mu)/T$\,, can be regarded as functions of $\epsilon$
only. The numerical solution of Eqs.~(\ref{emte1})--(\ref{emte2}) is
obtained by using the relativistic version~\cite{Ris95b} of the
flux--corrected transport algorithm~\cite{Bor73}. The details of the
computational procedure are given in Appendix~A.

\subsection{Initial conditions}

Following Ref.~\cite{Hir02}, we choose the initial conditions for a
finite-size fluid, generalizing the Bjorken scaling conditions:
\bel{incd}
Y (\tau_0,\eta)=\eta\,,\hspace*{5mm}\epsilon\hsp (\tau_0,\eta)=\epsilon_0
\exp{\left[-\frac{(|\eta|-\eta_0)^2}{2\sigma^2}
\Theta(|\eta|-\eta_0)\right]},
\ee
where $\Theta (x)\equiv (1+\textrm{sgn}\hsp x)/2$\,. All
calculations are performed for $\tau_0=1$\,fm/c, assuming that
$Y$ and $\epsilon$ are independent on the transverse coordinates.
The particular choice $\eta_0=0$ corresponds to the pure Gaussian
profile of the energy density. The case $\sigma=0$ corresponds
to the ''table--like'' profile
when the initial energy density equals $\epsilon_0$ at $|\eta|<\eta_0$
and vanishes at larger $|\eta|$\,. It is obvious that if $\sigma$ or
$\eta_0$ tends to infinity, one gets the limiting case of the Bjorken
scaling solution. In this case, as follows from
Eqs.~(\ref{bcce1})--(\ref{emte2}), $Y=\eta$ and $n,\epsilon,P$ are
functions of $\tau$, determined by the ordinary differential
equations~\cite{Bjo83}\footnote
{
By using thermodynamic relations one can derive from Eqs.~(\ref{bjoe})
the equation for entropy density:
\mbox{$d\hsp (s\tau)/d\tau=0$}.\label{sentr}
}
\bel{bjoe}
\frac{d\hsp (n\tau)}{d\tau}=0\,,~~~\frac{d\hsp (\epsilon\tau)}{d\tau}+P=0\,.
\ee
Comparison of initial energy density profiles corresponding to the same
total energy of produced particles, but having different
parameters~$\sigma$ and $\eta_0$, is made below in
Fig.~\ref{fig:eta-epsi}. The symmetry of central collisions of equal
nuclei with respect to the transformation $z\to -z$ implies that at
fixed $\tau$, including the initial state at $\tau=\tau_0$\,, the flow
rapidity and energy density should be, respectively, odd and even
functions of $\eta$\,:
\bel{parc}
Y\hsp (\tau,-\eta)=-Y\hsp (\tau,\eta)\,,~~~\epsilon\hsp (\tau,-\eta)=
\epsilon\hsp (\tau,\eta)\,.
\ee
Therefore, it is sufficient to solve Eqs.~(\ref{emte1})--(\ref{emte2})
only in the forward ''hemisphere'' $\eta\geqslant 0$\,, using the
boundary conditions $Y=0,\,\partial\epsilon/\partial\eta=0$ at
$\eta=0$\,.

\subsection{Equation of state\label{EOS}}

One of the main goals of experiments on ultrarelativistic heavy--ion
collisions is to study the deconfinement phase transition of
strongly--interacting matter. In our calculations this phase transition is
implemented through a bag--like
EOS using the parametrization suggested in Ref.~\cite{Tea01}. This EOS
consists of three parts, denoted below by indices $H, M, Q$\,, and
corresponding, respectively, to the hadronic, ''mixed'' and
quark--gluon phases. As already mentioned, pressure, temperature and
sound velocity, $c_s=\sqrt{dP/d\epsilon}$, of the baryon--free matter
can be regarded as functions of $\epsilon$\, only. It is further assumed
that $c_s$ is constant in each phase and, therefore, $P$ is a linear function of
$\epsilon$ with different slopes in the above--mentioned regions of
energy density.

The hadronic phase corresponds to the domain of low
energy densities, \mbox{$\epsilon<\epsilon_H$}, and temperatures,
$T<T_H$\hsp . This phase consists of pions, kaons, baryon--antibaryon pairs
and hadronic resonances. Numerical calculations for the ideal gas of
hadrons (see e.g.~\cite{Cho05}) predict a rather soft EOS:  the corresponding
sound velocity squared, $c_s^2=c_H^2\sim 0.1-0.2$\,, is noticeably lower than 1/3.
The mixed phase takes place at intermediate energy densities, from
$\epsilon_H$ to~$\epsilon_Q$ or at temperatures from~$T_H$
to~$T_Q$\,. The quantity $\epsilon_Q-\epsilon_H$ can be interpreted as
the ''latent heat'' of the deconfinement transition. To avoid numerical problems, we
choose a small, but nonzero value of sound velocity $c_M$ in the mixed
phase. The third, quark--gluon plasma region of the EOS corresponds to
$\epsilon>\epsilon_Q$ or $T>T_Q$. It is assumed that $c_s^2=c_Q^2$ reaches
the asymptotic value (1/3) already at the beginning of the quark--gluon
phase, i.e. at $T\simeq T_Q$\,.

The analytic expressions for pressure and temperature as functions of
$\epsilon$ in all three phases are given by the following
equations\footnote
{
Formally, to solve fluid--dynamical equations for the ideal
baryon--free matter, one needs only pressure as function of
$\epsilon$\,. However, to find particle momentum distributions, it is
necessary to know also temperature or entropy density of the fluid.
}
\begin{eqnarray}
\label{heos}
&&P=c_H^2\epsilon\,,~T=T_H\left(\frac{\epsilon}
{\epsilon_H}\right)^{\frac{\ds c_H^2}{\ds 1+c_H^2}}~~~(\epsilon<\epsilon_H)\,,\\
\label{meos}
&&P=c_M^2\epsilon-(1+c_M^2)\hsp B_M\,,~T=T_H\left(\frac{\epsilon-B_M}
{\epsilon_H-B_M}\right)^{\frac{\ds c_M^2}{\ds 1+c_M^2}}
~~~(\epsilon_H<\epsilon<\epsilon_Q)\,,\\
\label{qeos}
&&P=c_Q^2\epsilon-(1+c_Q^2)\hsp B_Q\,,~T=T_Q\left(\frac{\epsilon-B_Q}
{\epsilon_Q-B_Q}\right)^{\frac{\ds c_Q^2}{\ds 1+c_Q^2}}
~~~(\epsilon>\epsilon_Q)\,.
\end{eqnarray}
Here $B_M$ and $B_Q$ are the ''bag'' constants in the mixed and
quark phases, respectively. Our parameters
$c_H, c_M, c_Q$ are close to those used in Refs.~\cite{Kol99,Tea01}
The constants $B_M, B_Q$ and $T_Q$ are found from
\begin{table}[htb!]
\caption{Parameters of EOS\hsp s with the deconfinement phase transition.}
\label{tab1}
\bigskip
\hspace*{-4mm}
\begin{ruledtabular}
\begin{tabular}{l|c|c|c|c|c|c|c|c|c}
& $\epsilon_H$ & $\epsilon_Q$ &
$c_H^2$ & $c_M^2$ & $c_Q^2$ & $T_H$ & $T_Q$ &
 $B_M$ & $B_Q$ \\[-2mm]
& (GeV/fm$^3$) & (GeV/fm$^3$) &&&& (MeV) & (MeV) & (MeV/fm$^3$)
& (MeV/fm$^3$) \\
\hline
EOS--I & 0.45 & 1.65 &~0.15~&~0.02~&1/3~& 165 & 169 & -57.4 & 344\\
EOS--II & 0.79 & 2.90 &~0.15~&~0.02~&1/3~& 190 & 195 & -101 & 605\\
EOS--III& 0.45 & 1.65 & 1/3~&~0.02~&1/3~& 165 & 169 & -138 & 282\\[1mm]
\end{tabular}
\end{ruledtabular}
\end{table}
the continuity conditions for $P(\epsilon)$ and $T(\epsilon)$\,. The
corresponding formulas for the entropy density are obtained from the
thermodynamic relation $s=(\epsilon+P)/T$\,.
Below we use several versions of this EOS with the parameters
listed in Table~\ref{tab1}.
\begin{figure*}[htb!]
\centerline{\includegraphics[width=0.8\textwidth]{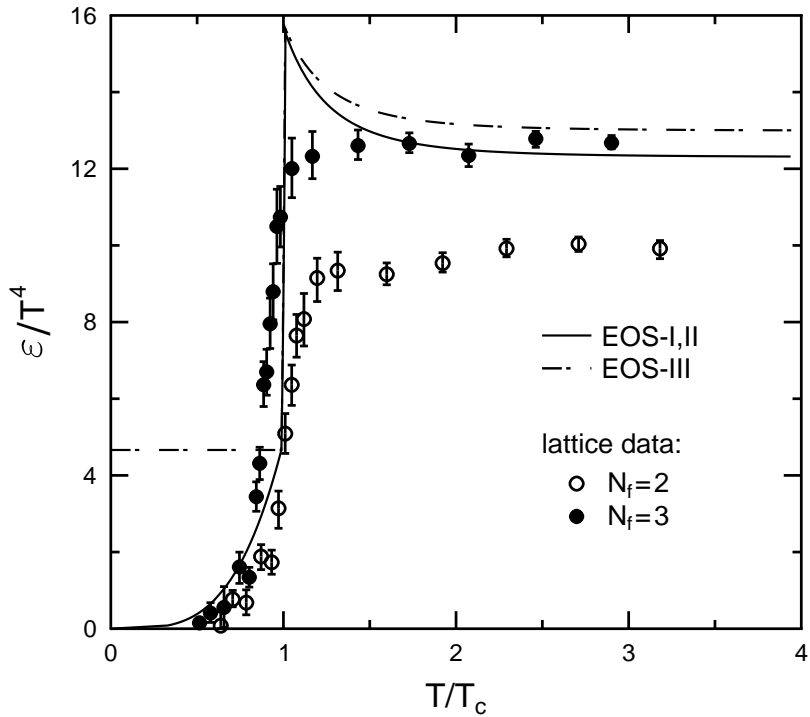}}
\caption{
Temperature dependence of energy density calculated from
Eqs.~(\ref{heos})--(\ref{qeos}) with parameters from Table~\ref{tab1}.
The solid line corresponds to both EOS--I ($T_c=167$\,MeV) and EOS--II
(\mbox{$T_c=192$\,MeV}). The dashed--dotted line is calculated for EOS--III
($T_c=167$\,MeV). Open (full) dots show the lattice data~\cite{Kar04}
for baryon--free matter with number of quark flavors $N_f=2\,(3)$\,.
}
\label{fig:eos-lat}
\end{figure*}

\begin{figure*}[htb!]
\vspace*{-1cm}
\centerline{\includegraphics[width=0.8\textwidth]{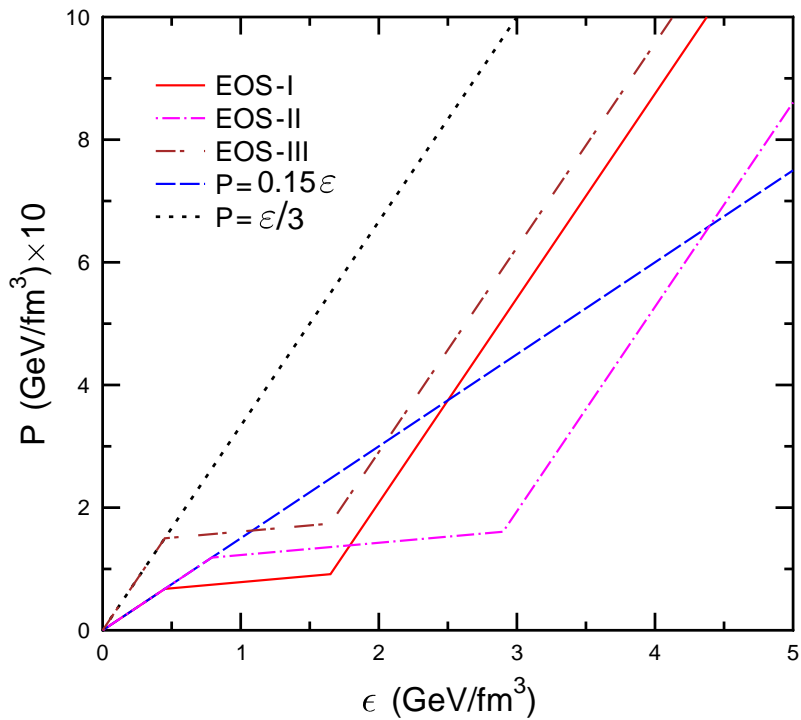}}
\caption{
Comparison of different EOSs used in this paper. The lines for EOS--I,
EOS--II and EOS--III are calculated using Eqs.~(\ref{heos})--(\ref{qeos})
with parameters from Table~\ref{tab1}.
}
\label{fig:eps-pres}
\end{figure*}
\begin{figure*}[htb!]
\centerline{\includegraphics[width=0.8\textwidth]{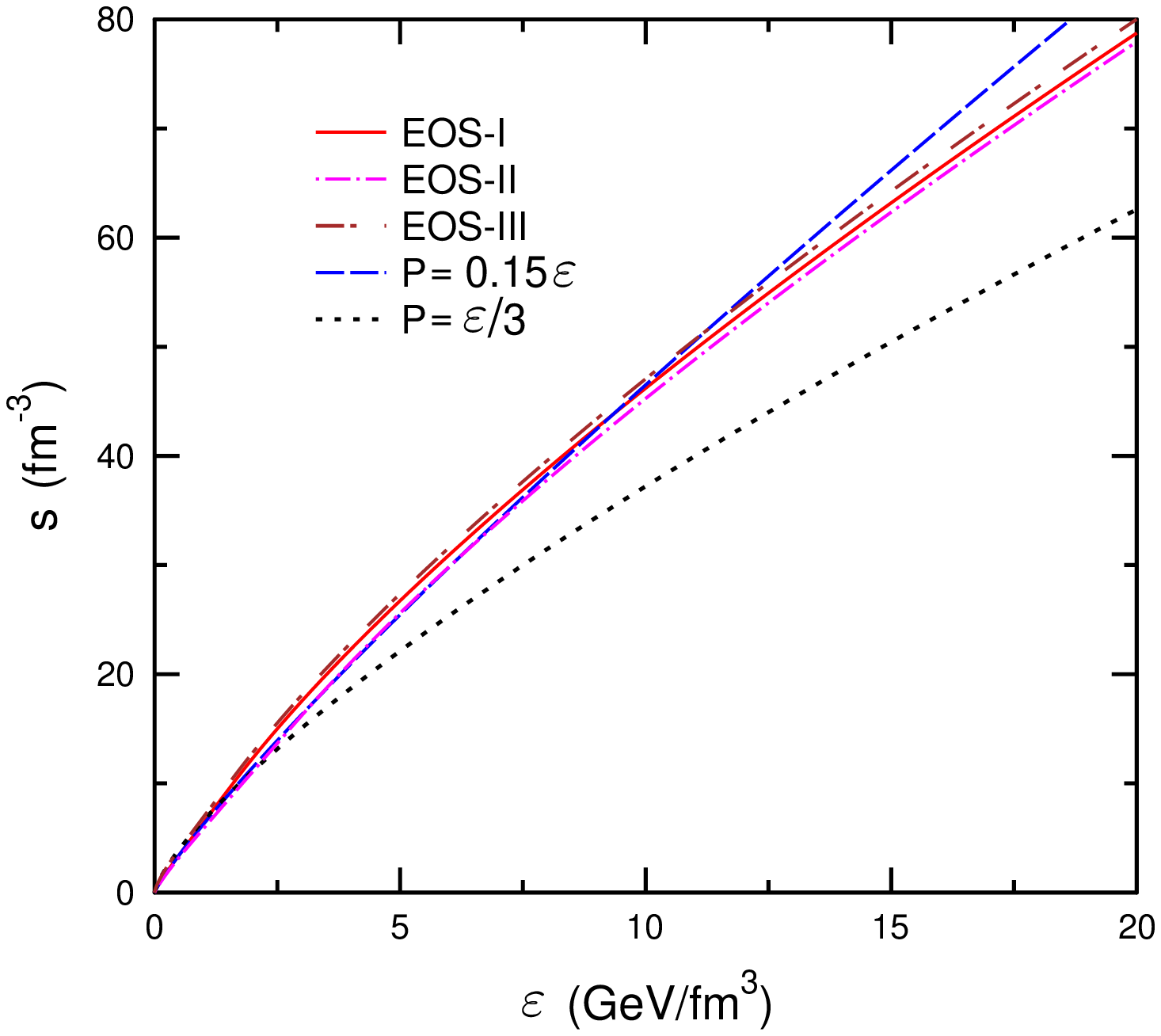}}
\caption{
Entropy density as a function of energy density for different EOS\hsp s.
}
\label{fig:eps-sden}
\end{figure*}

The parameters $T_H$ and $T_Q$ define the boundaries of a mixed phase
region separating the hadronic and quark--gluon phases. The critical
temperature~$T_c$ as defined by lattice calculations should lie between
$T_H$ and $T_Q$\,, i.e. $T_c\simeq (T_H+T_Q)/2$\,. Earlier lattice
calculations (see e.g.~Ref.~\cite{Kar01}) predicted the values
\mbox{$T_c=(170\pm 10)$\,MeV} for the baryon--free two--flavor QCD
matter. However, a noticeably larger value \mbox{$T_c=(192\pm
11)$\,MeV} has been reported recently in Ref.~\cite{Che06}. To probe
sensitivity to the actual position of the phase transition, we consider
the EOS\hsp s with different $T_H$ and $T_Q$\, (see Table~\ref{tab1}).
The EOS--I corresponds to $T_H=165$\,MeV and the parameters
$\epsilon_H,\,\epsilon_Q$ used in the parametrization LH12 of
Ref.~\cite{Tea01}\hsp\footnote
{
However, we choose a slightly lower value $c_H^2=0.15$.
}.
In the~\mbox{EOS--II} we choose $T_H=190$\,MeV and scale
$\epsilon_H,\,\epsilon_Q$ to get the same values of $\epsilon/T^4$ as a
function of~$T/T_H$\hsp\footnote
{
This is achieved by choosing the same $\epsilon_i/T_H^4\ (i=H,Q)$ for
these two EOSs.
}.
The EOS--III differs from the EOS--I by choosing a higher value
$c_H^2=1/3$\,.  Finally, the parameters $B_Q, T_Q$ are found from the
continuity conditions for $P$ and $T$\,.  As one can see from
Fig.~\ref{fig:eos-lat}, the parametrization (\ref{heos})--(\ref{qeos})
of the EOS leads to the cusp in $\epsilon/T^4$ at $T=T_c$\,,
typical for bag--like models. The lattice results for
$N_f=3$ are overestimated by about 30\%
in the temperature region from $T_c$ to $1.3\hsp T_c$\,.
An attempt to remove this discrepancy has been made in
Ref.~\cite{Moh03}.  It is also seen that the EOS--III strongly
disagrees with the lattice data in the hadronic phase at $T<T_c$\,.

Unless stated otherwise, these EOS\hsp s are used in the calculations
presented in this paper. For comparison, we have performed also
calculations for several purely hadronic EOS\hsp s. In this case we
extend~\re{heos} to energy densities $\epsilon>\epsilon_H$ with the
same $\epsilon_H, T_H$ as in Table~\ref{tab1}, but taking different
values of $c_H^2$ from 0.15 to 1/3. In Figs.~\ref{fig:eps-pres}--\ref{fig:eps-sden}
we compare our EOS\hsp s with the phase transition and two hadronic
EOS\hsp s with constant sound velocities $c_s=c_H$\,. According to
Figs.~\ref{fig:eps-pres}, the mixed phase region in the EOS--II
occupies a larger interval of energy densities, i.e. this EOS has a
larger latent heat as compared to the EOS--I and EOS--III. By this
reason, the life--time of the mixed phase should be longer for the
EOS--II, assuming the same initial state.  One can see that the
hadronic EOS $P=c_H^2\epsilon$ with $c_H^2=0.15$ (the dashed line in
Fig.~\ref{fig:eps-pres}) is much softer at high energy densities as
compared to the EOS--I and EOS--II. However, according to
Fig.~\ref{fig:eps-sden}, the difference between entropy densities
predicted by this hadronic EOS and the EOS\hsp s with pase transition
is rather small at $\epsilon\lesssim 10$ GeV/fm$^{3}$. Due to these
reasons, the rapidity spectra of pions and kaons predicted by our model
for soft EOS\hsp s, corresponding to $c_H^2\lesssim 0.2$, are, in
general, rather insensitive to the deconfinement phase transition.  On
the other hand, the EOS of massless particles, $P=\epsilon/3$\,, differs
significantly from other EOS\hsp s shown in Fig.~\ref{fig:eps-sden}.
As a consequence, visible differences with purely hadronic calculations
appear only for hard EOS\hsp s, i.e. for $c_H^2\gtrsim 1/3$\,.

\subsection{Total energy and entropy of the fluid\label{sum_rules}}

The equations of ideal fluid dynamics (\ref{bcce})--(\ref{enmt}) imply
that the total baryon number $B$, total 4--momentum $P^{\hsp\mu}$ and
entropy $S$ are the same at any hypersurface $\sigma^\mu$ lying above
the initial hypersurface (in our case $\tau=\tau_0$):
\begin{eqnarray}
\label{tbarn0}
&&B=\int\hspace*{-1mm} d\sigma_\nu n\hsp U^{\nu}=\textrm{const}\,,\\
\label{tfmom}
&&P^\mu=\int\hspace*{-1mm} d\sigma_\nu T^{\mu\nu}=\textrm{const}\,,\\
\label{tentr0}
&&S=\int\hspace*{-1mm} d\sigma_\nu s\hsp U^{\nu}=\textrm{const}\,.
\end{eqnarray}
In the case of central collision of identical nuclei, due to symmetry
reasons, the total three--momentum is zero in the c.m. frame.
Therefore, Eqs.~(\ref{tfmom}) give only one nontrivial constraint,
namely, the conservation of the total c.m. energy, $E=P^0$\,.

Let us consider a cylindrical volume of matter (''fireball'') expanding
in the longitudinal direction and choose the hypersurface of constant
$\tau$\,. The elements of this hypersurface may be written as
\bel{hypel}
d\sigma^\mu=S_\perp (dz,\bm{0},dt)^\mu=\tau S_\perp d\eta\hsp (\cosh{\eta},
\bm{0},\sinh{\eta})^\mu,
\ee
where $S_\perp=\pi R^2$ is the transverse cross section of the
fireball. Substituting (\ref{hypel}) into Eqs.~(\ref{tbarn0})--(\ref{tentr0})
we get the relations
\begin{eqnarray}
\label{tbarn}
&&B=S_\perp\tau\int\limits_{-\infty}^{+\infty}
\hspace*{-1mm}d\eta\,n\cosh{(Y-\eta)}\,,\\
\label{tener}
&&E=S_\perp\tau\int\limits_{-\infty}^{+\infty}\hspace*{-1mm}d\eta
\left[\hsp\epsilon\hsp\cosh{Y}\cosh{(\mbox{$Y-\eta$})}+
P\sinh{Y}\sinh{(Y-\eta)}\hsp\right]\hspace*{-1pt},\\
\label{tentr}
&&S=S_\perp\tau\int\limits_{-\infty}^{+\infty}
\hspace*{-1mm}d\eta\,s\cosh{(Y-\eta)}\,.
\end{eqnarray}
for any fixed $\tau\geqslant\tau_0$\,. One can determine constants in
Eqs.~(\ref{tbarn0})--(\ref{tentr0}) by using Eqs.~\mbox{(\ref{tbarn})--(\ref{tentr})}
for~\mbox{$\tau=\tau_0$}. In the baryon--free case ($n$ and $B$ are
zero), substituting $\tau=\tau_0,\, Y=\eta$
into~\mbox{(\ref{tener})--(\ref{tentr})} leads to
\begin{eqnarray}
\label{tener1}
&&E=2\hsp S_\perp\tau_0\int\limits_0^\infty\hspace*{-1mm}d\eta\,
\epsilon(\tau_0,\eta)\hsp\cosh{\eta}\,,\\
\label{tentr1}
&&S=2\hsp S_\perp\tau_0\int\limits_0^\infty
\hspace*{-1mm}d\eta\,s(\tau_0,\eta)\,.
\end{eqnarray}
Equations (\ref{tener1})--(\ref{tentr1}) with $E$ and $S$ from
(\ref{tener})--(\ref{tentr}) can be considered as two sum rules for the
fluid--dynamical quantities. We have checked that our numerical code
conserves the total energy and entropy at any hypersurface $\tau={\rm
const}$ on the level better than~1\% even for \mbox{$\tau\gtrsim
200$\,fm/c}.

\begin{figure*}[htb!]
\centerline{\includegraphics[width=0.8\textwidth]{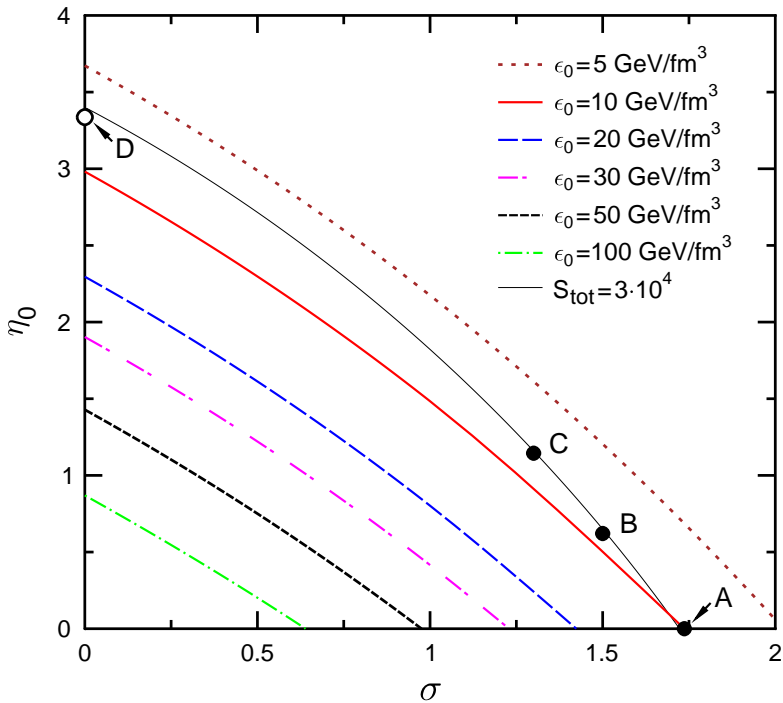}}
\caption{
Contours of constant $\epsilon_0$ in the $\sigma-\eta_0$ plane,
corresponding to the fixed total energy~(\ref{teval}) of initial
fireball in central Au+Au collision ($R=6.5$\,fm). Points A--D
represent the parameter sets included into the detailed analysis (see
below).  Point A (D) corresponds to the Gaussian (table--like) initial
condition.  Thin solid line shows the contour of total entropy
$S=3\cdot 10^4$\,, calculated for the EOS--I.
}
\label{fig:sig-eta0}
\end{figure*}
Below we use~\re{tener1} to constrain possible values of the parameters
characterizing the initial state. This is possible since the total
energy of produced particles can be estimated from experimental data.
Indeed, the value of the total energy loss, $\Delta E=73\pm 6$ GeV per
participating nucleon, has been obtained from the net baryon rapidity
distribution in most central Au+Au collisions~\cite{Bea04}. This gives
an estimate of the total energy of secondaries in the considered
reaction:
\bel{teval}
E=N_{\rm part}\hsp\hsp\Delta E\simeq 26.1\,\textrm{TeV}\,,
\ee
where \mbox{$N_{\rm part}\simeq 357$} is the mean number of
participating nucleons.

\begin{figure*}[htb!]
\centerline{\includegraphics[width=0.8\textwidth]{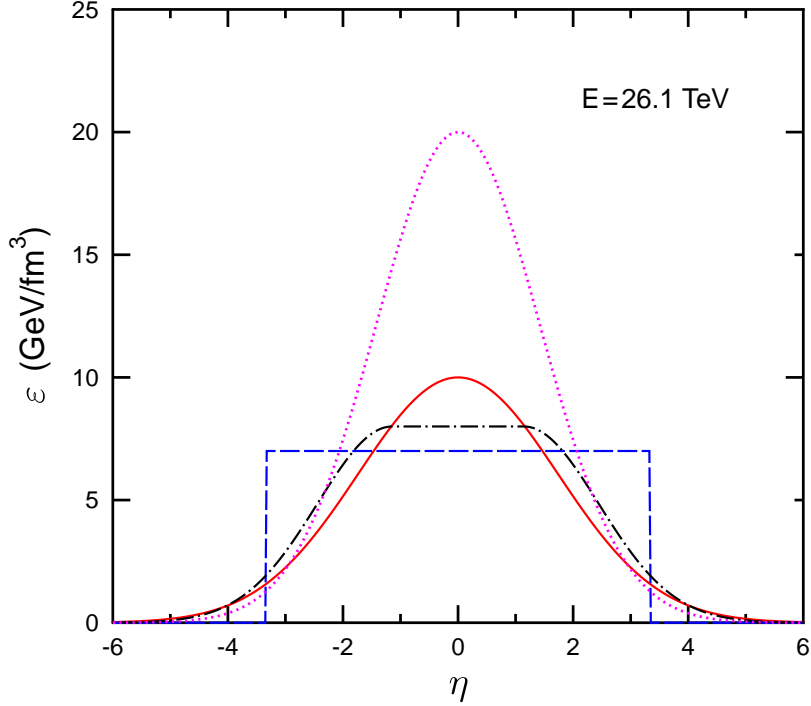}}
\caption{
Initial energy density profiles corresponding to the total energy
$E=26.1$\,TeV in central Au+Au collisions. Solid and dashed--dotted
lines correspond, respectively, to the parameter sets A and~C (see
Table~\ref{tab3}). Dashed line is calculated for the table--like
profile with $\eta_0=3.34$\,, $\epsilon_0=7$\,GeV/fm$^3$ (point D in
Fig.~\ref{fig:sig-eta0}). Dotted line corresponds to the Gaussian
profile with $\sigma=1.42$\,, $\epsilon_0=20$\,GeV/fm$^3$\,.
}
\label{fig:eta-epsi}
\end{figure*}
Substituting the parametri\-zation (\ref{incd}) into~\re{tener1} and
taking the value of $E$ from~\re{teval}, one gets the relation between
the parameters~$\epsilon_0, \eta_0, \sigma$. The integration in the
r.h.s.~of \re{tener} can be performed analytically. This gives
\bel{tener2}
\lambda=\alpha(\sigma)\cosh{\eta_0}+\beta(\sigma)\sinh{\eta_0}\,.
\ee
Here $\lambda=E/(2S_\perp\epsilon_0\tau_0)$ and $\alpha, \beta$ are
determined as
\bel{albe}
\alpha=\sqrt{\frac{\pi}{2}}\,\sigma\exp{\left(\sigma^2/2\right)}\,,
~~~\beta=1+\alpha\hsp{\rm erf}\left(\frac{\sigma}{\sqrt{2}}\right).
\ee
It is easy to see that the solution of \re{tener2} with respect to
$\eta_0$ exists at $\alpha<\lambda$\,, i.e. at not too large $\sigma$.
This solution is given by the expression
\bel{etas}
\eta_0=\ln{\left(\frac{\ds \lambda+\sqrt{\lambda^2+\beta^2-\alpha^2}}
{\alpha+\beta}\right)}\,.
\ee

\begin{figure*}[htb!]
\centerline{\includegraphics[width=0.8\textwidth]{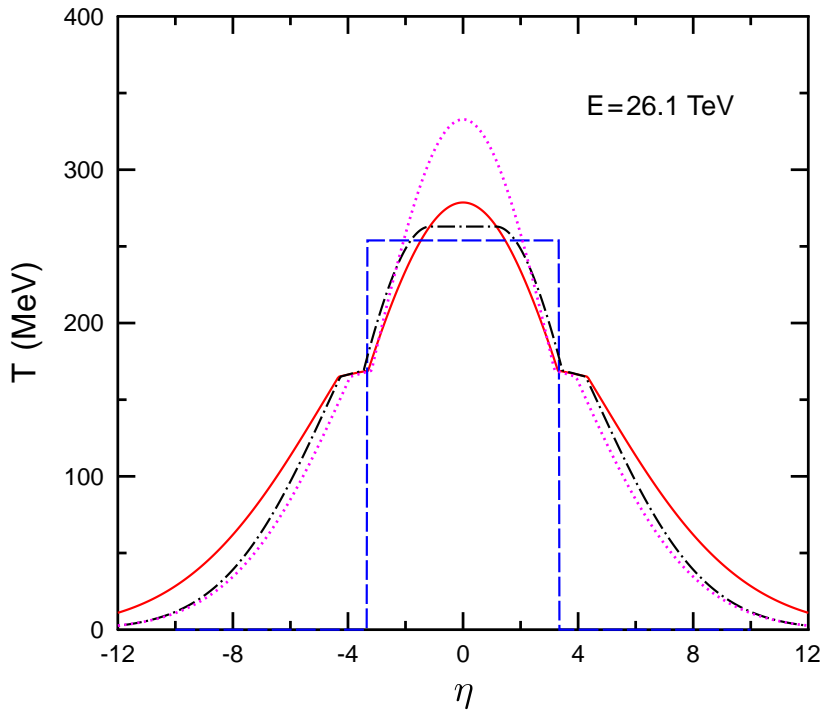}}
\caption{
Same as Fig.~\ref{fig:eta-epsi}, but for initial temperature profiles.
All curves correspond to the EOS--I.}
\label{fig:eta-temi}
\end{figure*}
Figure~\ref{fig:sig-eta0} shows contours of constant $\epsilon_0$ in
the $\eta_0-\sigma$ plane. All curves are calculated for the same total
energy of the fluid given in~\re{teval}. The points at the horizontal
(vertical) axis correspond to Gaussian (table--like) initial energy
density profiles. One can see that these profiles become narrower for
higher maximal energy densities $\epsilon_0$  (in this case the
contours of equal $\epsilon_0$ move closer to the origin of the
$\sigma-\eta_0$ plane). It is obvious that an~additional constraint on the
total entropy $S$ would further reduce the freedom in the choice of
parameters $\epsilon_0, \eta_0, \sigma$\,. For example, the best fits
of the BRAHMS data (see below) with the EOS--I fall on the line corresponding to
$S\simeq 3\cdot 10^4$\,.
According to our analysis, the absolute yields of secondary mesons and
antibaryons increase with $S$, even if the total energy is fixed. This
effect is well--known from the Landau model~\cite{Lan53} which postulates
that the number of secondaries is proportional to $S$\,. The initial
temperatures for the considered profiles reach the values 250--350 MeV,
well above the critical temperature of the quark--gluon phase transition.

Figure \ref{fig:eta-epsi} represents some profiles of initial energy
density used in this paper. All these profiles correspond to the same
total energy of secondaries~(\ref{teval}). The corresponding profiles
of the initial temperature, $T\hsp (\tau_0,\eta)$, are shown in
Fig.~\ref{fig:eta-temi}. They are calculated by using the EOS--I. Note,
that at nonzero~$\sigma$ the temperature profiles look much wider as
compared to those shown in Fig.~\ref{fig:eta-epsi}. One can see that
three different phases of matter appear already at the initial state.

\subsection{Particle spectra at freeze--out\label{sfreez}}

The momentum spectra of secondary hadrons are calculated by applying
the standard Cooper--Frye formula~\cite{Coo74}, assuming that particles
are emitted without further rescatterings from the elements
$d\sigma_\mu$ of the freeze--out hypersurface $\tau=\tau_F(\eta)$\hsp .
Then, the invariant momentum distribution for each particle species is
given by the expression
\bel{spec}
{E}\frac{d^{\hsp 3}\hspace*{-1pt}N}{\hspace*{-4pt}d^{\hsp 3}p}=
\frac{d^{\hsp 3}\hspace*{-1pt}N}{dy\hsp d^{\hsp 2}p_T}=\frac{g}{(2\pi)^3}
\int d\sigma_\mu\hsp p^{\hsp\mu}\left\{\exp\left(\frac{p\hsp U_F-\mu_F}
{T_F}\right)\pm 1\right\}^{-1},
\ee
where \mbox{$p^{\hsp\mu}=(E,\bm{p})^\mu$} is the 4--momentum of the
particle, \mbox{$y=\tanh^{-1}(p_z/E)$} and $\bm{p}_T$ are,
respectively, its longitudinal rapidity and transverse momentum, $g$
denotes the particle's statistical weight. The subscript $F$ in the
collective 4--velocity $U$\,, temperature $T$ and chemical potential
$\mu$ implies that these quantities are taken on the freeze--out
hypersurface\footnote
{
Below we assume that the chemical and thermal freeze--out hypersurfaces
coincide. In this case $\mu_F=0$ for baryon--free matter.
}.
The plus or minus sign in the r.h.s. of~\re{spec} correspond to fermions or
bosons, respectively.

As has been already stated, the effects of transverse expansion are
disregarded in our approach. Due to this reason, we cannot describe
realistically the $p_T$ spectra of produced hadrons, and analyze below
only the rapidity spectra. For a cylindrical fireball with transverse
cross section $S_\perp=\pi R^2$ expanding only in the longitudinal
direction, one can write
$d\sigma^\mu=S_\perp\hsp (dz_F, \bm{0}, dt_F)^{\hsp\mu}$\,.
Using~\re{lcva} we get the following relations
\bel{free}
d\sigma_\mu p^{\hsp\mu}=S_\perp\hsp (E\hsp dz_F-p_z\hsp dt_F)=
S_\perp\hsp m_T \left\{\tau_F (\eta)\cosh\hsp (y-\eta)-
\tau^\prime_F (\eta)\sinh\hsp (y-\eta)\right\} d\eta\,.
\ee
Here $m_T$ is the particle's transverse mass defined as
$m_T=\sqrt{m^2+\bm{p}_T^{\hsp 2}}$\,, where~$m$ is the corresponding
vacuum mass. Substituting (\ref{free}) into (\ref{spec}) and
integrating over $\bm{p}_T$ we get the expression for the particle
rapidity distribution
\begin{eqnarray}
\nonumber
&&\frac{dN}{dy}=\frac{g S_\perp}{4\pi^2}\int\limits_{-\infty}^{+\infty}\hspace*{-1mm}
d\eta\left[\tau_F (\eta)\cosh\hsp (y-\eta)
-\tau^\prime_F (\eta)\sinh\hsp (y-\eta)\right]\times\\
&&\times\int\limits_{0}^{\infty}\hspace*{-1mm}dp_T p_T\hsp m_T
\left\{\exp\left(\frac{m_T\cosh\hspace*{-1pt}\left[\hsp y-Y_F(\eta)\right]-\mu_F}
{T_F(\eta)}\right)\pm 1\right\}^{-1},
\label{ydis}
\end{eqnarray}
where $Y_F(\eta)$ and $T_F(\eta)$ are the flow rapidity and the
temperature at $\tau=\tau_F(\eta)$\,. Note, that the Bjorken
model~\cite{Bjo83} corresponds to $Y_F=\eta$ and $\tau_F, T_F$
independent of $\eta$. As can be seen from~\re{ydis}, the rapidity
distributions of all particles should be flat in this case.

\subsection{Feeding from resonance decays}

In calculating particle spectra one should take into account not only
directly produced particles but also feeding from resonance decays. In
this paper we explicitly take into account the decays:
$\rho\to 2\pi, K^*\to K\pi, \ov{\Delta}\to\ov{N}\pi$\,,
which are most important for describing the spectra of pions, kaons and
antiprotons, respectively. The other resonances are taken into account
using an approximate procedure explained below. The spectrum of
$i$--th particle produced in the two particle decay channel $R\to iX$
is calculated by using the expression~\cite{Sol90}
\bel{rdc}
E\frac{d^{\hsp 3}N_{R\to i}}{d^{\hsp 3} p\hspace*{3pt}}=
\frac{d^{\hsp 3}N_{R\to i}}{d^{\hsp 2}p_T dy}=
\int\limits_{\ds m_{\rm thr}}^\infty d\hsp m_R
\hsp w\hsp (m_R)\int d^{\hsp 3}p_R\hsp\frac{d^{\hsp 3}N_R}{d^{\hsp 3}p_R}\cdot
\frac{b_{R\to i}}{4\pi q_0}\,
\delta\hspace*{-2pt}\left(\frac{p\hsp p_R}{m_R}-E_{0\hsp i}\right),
\ee
where the first integration corresponds to averaging over the mass
spectrum of the resonance ($m_{\rm thr}=m_i+m_X$),
$p_R=(E_R,\bm{p}_R)$~and $p=(E,\bm{p})$ are, respectively, the
4--momenta of the resonance $R$ and the particle $i$\,.  The
coefficient $b_{R\to i}$ denotes the branching ratio of the $R\to iX$
decay. The energy and momentum of the i--th particle in the resonance
rest frame are determined by the expressions
\bel{eprd}
E_{0\hsp i}=\sqrt{m_i^2+q_0^2}=\frac{m_R^2+m_i^2-m_X^2}{2m_R}\,.
\ee
The freeze--out momentum spectrum of the resonance,
$d^{\hsp 3}N_R/d^{\hsp 3}p_R$\,, is calculated using
Eqs.~(\ref{spec})--(\ref{free}) with \mbox{$m=m_R$}, \mbox{$g=g_R$}.
Below we assume that the freeze--out temperatures for directly produced
particles and corresponding resonances are the same. Calculation of
integrals in the r.h.s. of~\re{rdc} and the parametrization of $w\hsp
(m_R)$ are discussed in Appendix~B.
\begin{table}[htb!]
\caption{
Equilibrium densities of $\pi^+$\hsp --, $K^+$\hsp--mesons and
antiprotons hidden, respectively, in $\rho, K^*$ and $\ov{\Delta}$
resonances as well as corresponding enhancement factors at different
temperatures~$T_F$.}
\label{tab2}
\bigskip
\begin{ruledtabular}
\begin{tabular}{c|c|c|c|c|c|c}
~$T_F$\,(MeV)&~$d_\rho^{\hsp\pi^+} n_\rho$\,(fm$^{-3}$) &
$\alpha_\rho^{\hsp\pi^+}$ &~$d_{K^*}^{K^+} n_{K^*}$\,(fm$^{-3}$)&
~$\alpha_{K^*}^{K^+}$ &
~$d_{\hsp\ov{\Delta}}^{\hsp\ov{p}} n_{\ov{\Delta}}$\,(fm$^{-3}$)&
$\alpha_{\hsp\ov{\Delta}}^{\hsp\ov{p}}$\\
\hline
100 & $5.74\cdot 10^{-4}$ &~2.30 & $1.08\cdot 10^{-4}$ &~1.21 &
$1.48\cdot 10^{-5}$ &~1.09\\
110 & $1.37\cdot 10^{-3}$ &~2.28 & $2.85\cdot 10^{-4}$ &~1.26 &
$5.30\cdot 10^{-5}$ &~1.12\\
120 & $2.87\cdot 10^{-3}$ &~2.31 & $6.50\cdot 10^{-4}$ &~1.30&
$1.56\cdot 10^{-4}$ &~1.15\\
130 & $5.43\cdot 10^{-3}$ &~2.37 & $1.32\cdot 10^{-3}$ &~1.35 &
$3.92\cdot 10^{-4}$ &~1.18\\
140 & $9.48\cdot 10^{-3}$ & ~2.46 & $2.46\cdot 10^{-3}$ &~1.40 &
$8.74\cdot 10^{-4}$ &~1.21\\
150 & $1.55\cdot 10^{-2}$ &~2.57 & $4.24\cdot 10^{-3}$ &~1.46 &
$1.77\cdot 10^{-3}$ &~1.24\\
160 & $2.41\cdot 10^{-2}$ &~2.70 & $6.90\cdot 10^{-3}$ &~1.51 &
$3.29\cdot 10^{-3}$ &~1.28\\
165 & $2.95\cdot 10^{-2}$ &~2.77 & $8.63\cdot 10^{-3}$ &~1.54 &
$4.39\cdot 10^{-3}$ &~1.29\\
170 & $3.58\cdot 10^{-2}$ &~2.83 & $1.07\cdot 10^{-2}$ &~1.57 &
$5.75\cdot 10^{-3}$ &~1.31\\
180 & $5.13\cdot 10^{-2}$ &~2.98 & $1.58\cdot 10^{-2}$ &~1.62 &
$9.50\cdot 10^{-3}$ &~1.34\\
190 & $7.11\cdot 10^{-2}$ &~3.13 & $2.27\cdot 10^{-2}$ &~1.68 &
$1.50\cdot 10^{-2}$ &~1.37\\[1mm]
\end{tabular}
\end{ruledtabular}
\end{table}

\begin{figure*}[htb!]
\centerline{\includegraphics[width=0.8\textwidth]{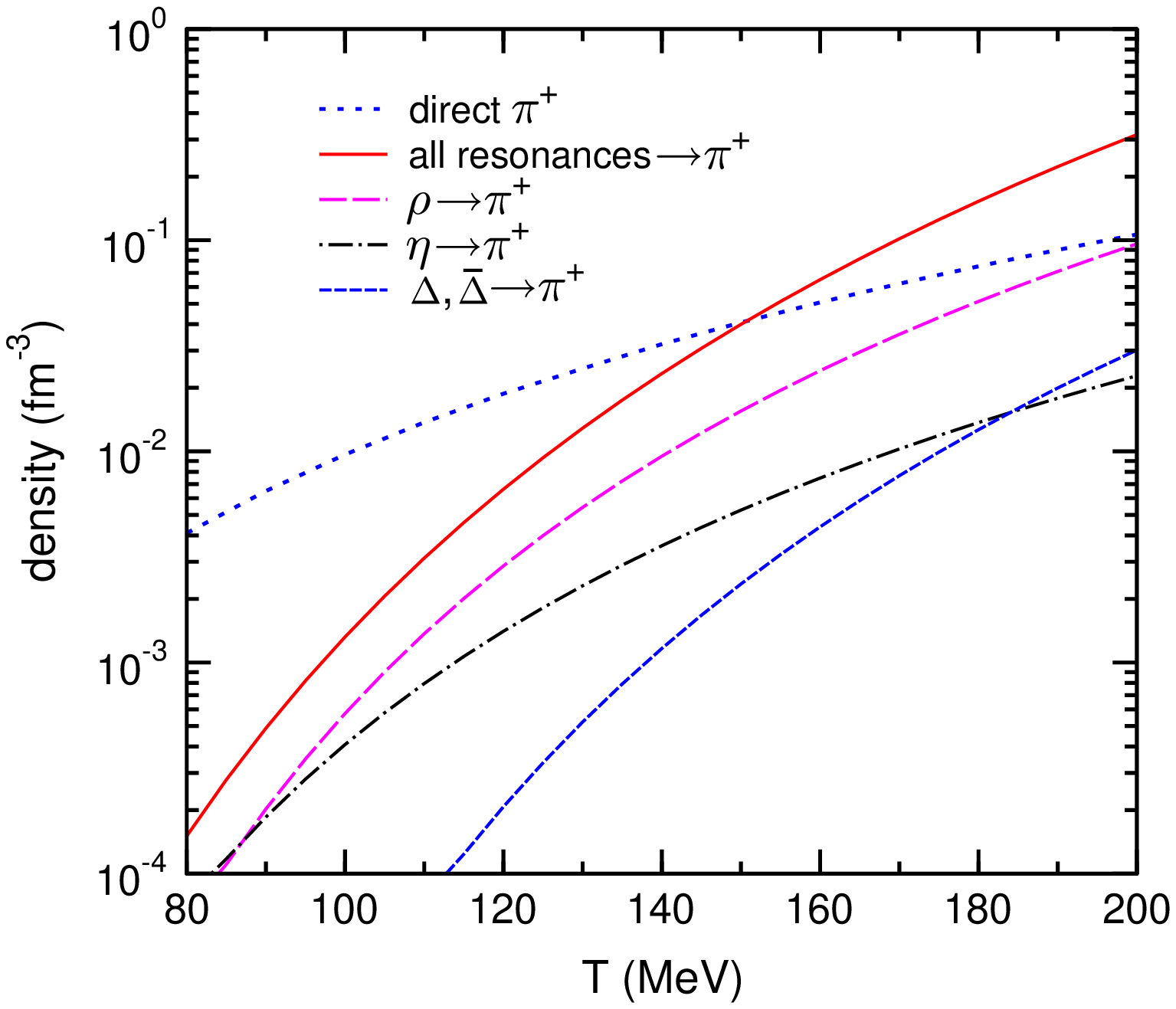}}
\caption{
Equilibrium densities of $\pi^+$ mesons hidden in resonances
$\eta,\rho,\Delta,\ov{\Delta}$ as functions of temperature $T$\,.
Solid line shows sum over all mesonic and baryonic resonances (see
text). Dotted line represents the equilibrium density of free
$\pi^+$--mesons (not hidden in resonances).
}
\label{fig:tem-den}
\end{figure*}
The detailed calculations in this paper are made for spectra of charged
pions ($i=\pi^+$), kaons ($i=K^+$) and antiprotons ($i=\ov{p}$). In
these cases the most important contributions are given by decays of
resonances $R=R_0=\rho,\,K^*(892),\,\ov{\Delta}\hsp (1232)$\,,
respectively.  To take into account other hadronic resonances
($R\neq R_0$) we assume that the contribution of the resonance~$R$ is
proportional to its equilibrium density, $n_R\hsp (T_F)$
multiplied by $d_R^{\,i}$, the average number of i--th particles
produced in  $R\to iX$ decays (e.g.
$d_\rho^{\,i}=2/3, \mbox{$d_\eta^{\,i}=0.65$}\ldots$\,\,for $i=\pi^+$).
Within this approximation one has
\bel{resc}
\sum\limits_R\frac{\ds d^{\hsp 3}N_{R\to i}}{\ds dy\hsp d^{\hsp 2}p_T}=
\alpha_{R_0}^i\hsp\frac{\ds d^{\hsp 3}N_{R_0\to i}}
{\ds dy\hsp d^{\hsp 2}p_T}\,,
\ee
where the enhancement factor $\alpha_{R_0}^{i}$ is
\bel{enf}
\alpha_{R_0}^i=\sum\limits_R\frac{d_R^{\hsp i}}{d_{R_0}^{\,i}}
\frac{n_R\hsp (T_F)}{n_{R_0}(T_F)}\,.
\ee
Here $n_R$ is the total density of resonances $R$ at the temperature $T$
summed over all isospin states of $R$\,. For simplicity, we calculate
$n_R$ in the zero width approximation. More details of $n_R$ and $d_R^{\,i}$
calculations can be found in Ref.~\cite{Beb92}. We have
checked for several resonances with
two--body decays (e.g. for $i=\pi$ and $R=K^*$) that such a procedure
yields a very good accuracy.  We include meson (baryon and antibaryon)
resonances with masses up to 1.3 (1.65) GeV and widths $\Gamma<150$ MeV.
The statistical weights, masses and branching ratios of these
resonances are taken from Ref.~\cite{PDG04}\,.  The results for
different $T_F$ are shown in Table~\ref{tab2}.  One can see that
factors $\alpha_R^i$ significantly exceed unity (especially for pions)
and increase with temperature.

Figure~\ref{fig:tem-den} shows the temperature dependence of
$\pi^+$--meson densities, $d_R^{\hsp\pi^+} n_R$, hidden in different
resonances $R$\,. One can see that, compared to other resonance
decays, the channels $\rho\to 2\pi$ give the largest contribution at
all temperatures. According to this calculation, in equilibrium
hadronic matter the fraction of~''free'' pions (not hidden in
resonances) is smaller than 50\% at temperatures $T\gtrsim
150$\,MeV.

\section{Results}\label{sres}

\subsection{Dynamical evolution of matter\label{devm}}

\begin{figure*}[htb!]
\hspace*{-15mm}\includegraphics[width=0.9\textwidth]{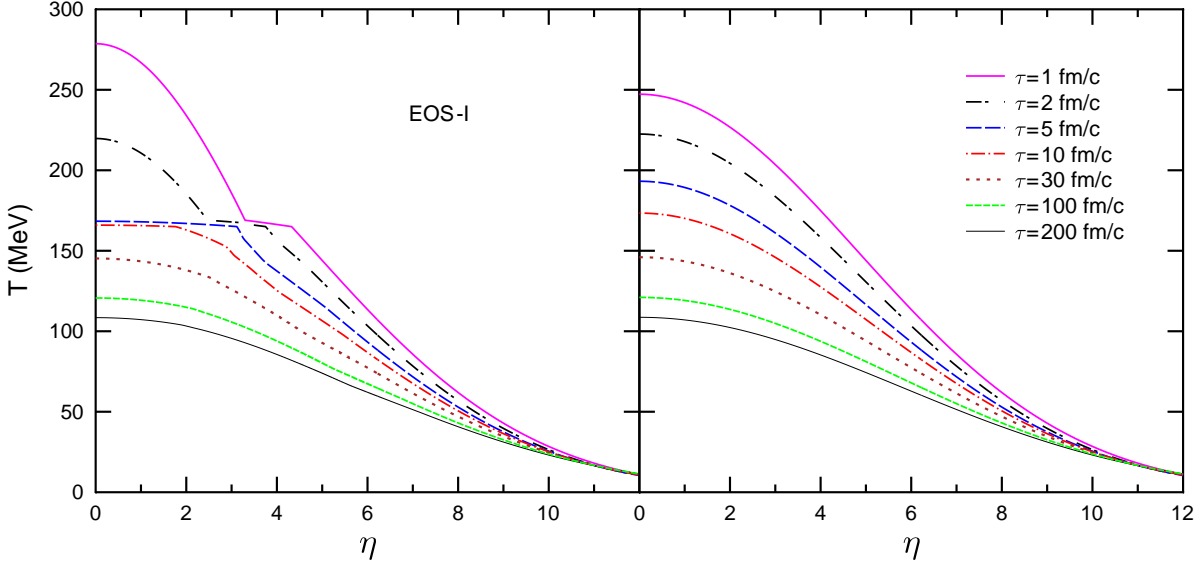}
\caption{
Temperature profiles at different proper times $\tau$ calculated for
initial conditions corresponding to the parameter set A.
Only the forward hemisphere ($\eta\geqslant 0$) is shown. Left and
right panels correspond, respectively, to the EOS--I and the hadronic
EOS $P=c_H^2\hsp\epsilon$\, with $c_H^2=0.15$\,.
}
\label{fig:eta-tem1}
\end{figure*}
\begin{figure*}[htb!]
\hspace*{-15mm}\includegraphics[width=0.9\textwidth]{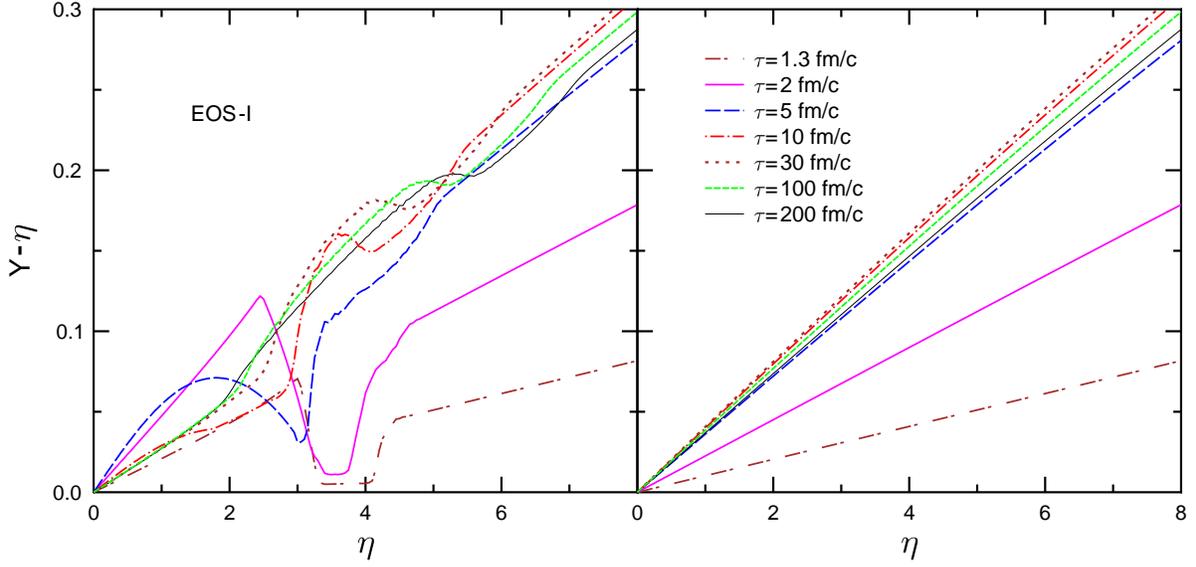}
\caption{
Same as Fig.~\ref{fig:eta-tem1}, but for collective rapidity profiles.
}
\label{fig:eta-y1}
\end{figure*}
\begin{figure*}[htb!]
\hspace*{-15mm}\includegraphics[width=0.9\textwidth]{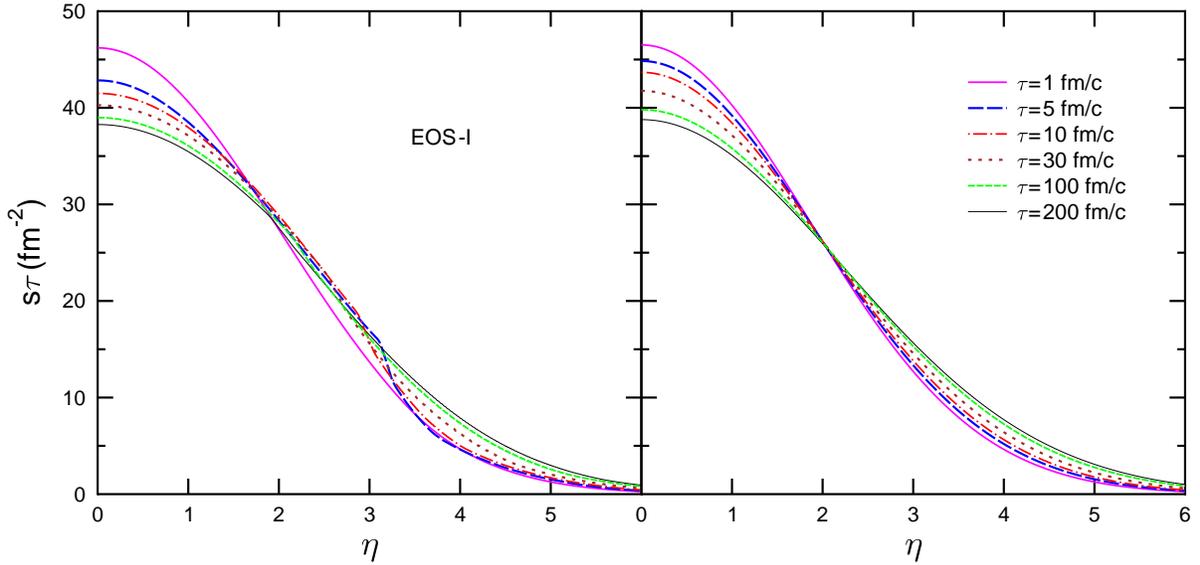}
\caption{
Same as Fig.~\ref{fig:eta-tem1}, but for profiles of entropy density
multiplied by $\tau$\,.
}
\label{fig:eta-stau}
\end{figure*}
\begin{figure*}[htb!]
\hspace*{-15mm}\includegraphics[width=0.9\textwidth]{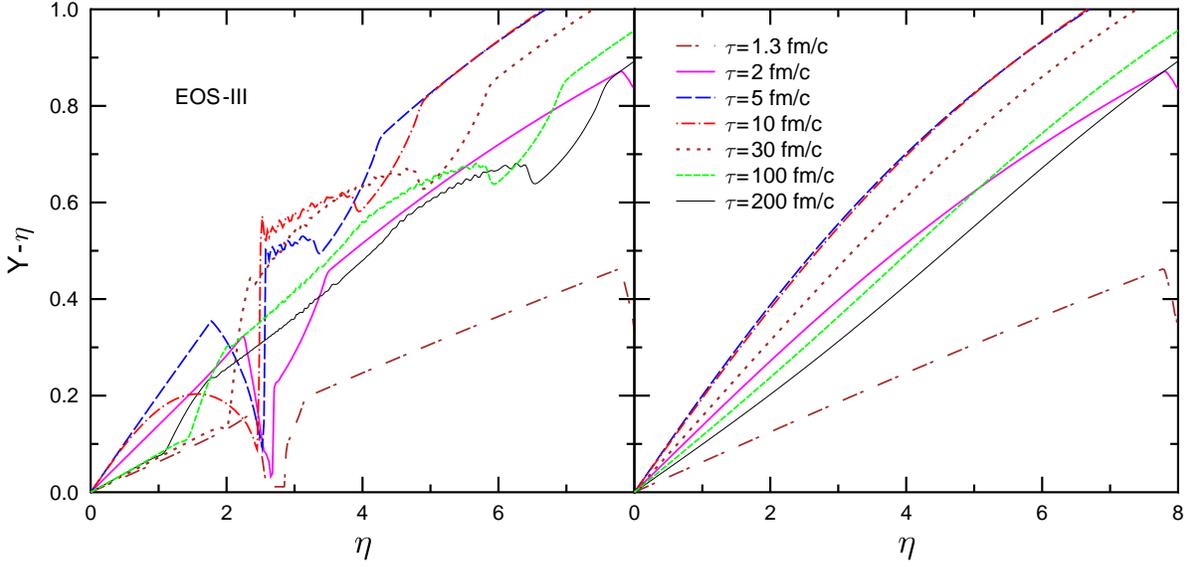}
\caption{
Collective rapidity profiles at different proper times $\tau$
calculated for initial conditions with parameters
$\epsilon_0=50$\,GeV/fm$^3,\,\eta_0=0,\,\sigma=0.98$\,.
Left and right panel correspond, respectively, to the EOS--III
and the hadronic EOS $P=\epsilon/3$\,.
}
\label{fig:eta-y2}
\end{figure*}
\begin{figure*}[htb!]
\hspace*{-15mm}\includegraphics[width=0.9\textwidth]{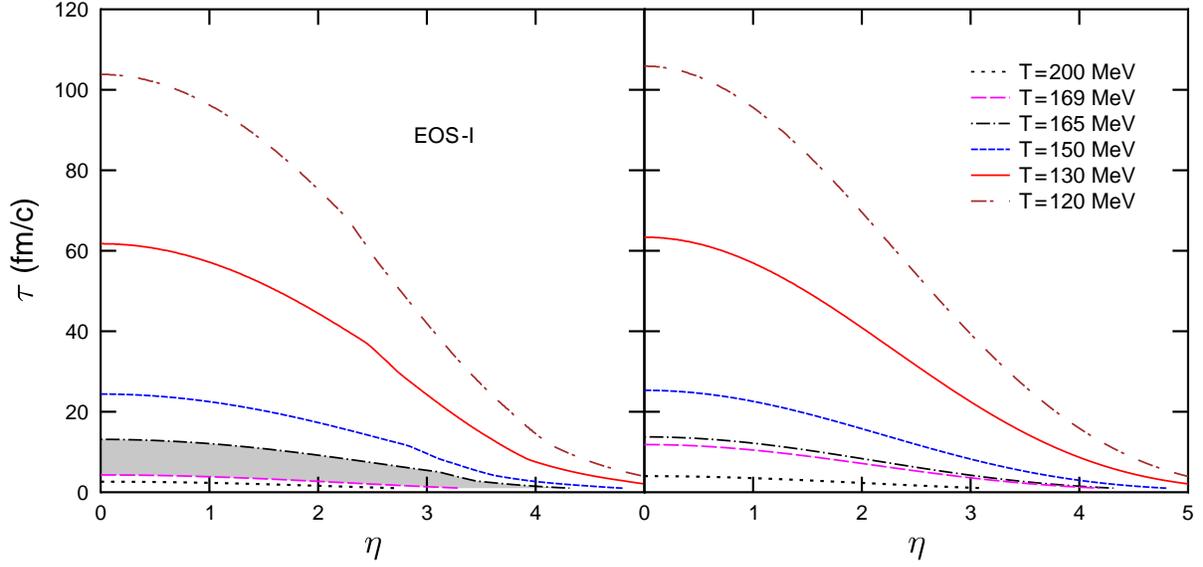}
\caption{
Isotherms in the $\eta-\tau$ plane calculated for the parameter set A.
Left and right panels correspond to the same EOS\hsp s as
in Fig.~\ref{fig:eta-tem1}. Shaded region indicates the mixed phase.
}
\label{fig:eta-tauf}
\end{figure*}
\begin{figure*}[htb!]
\hspace*{-15mm}\includegraphics[width=0.9\textwidth]{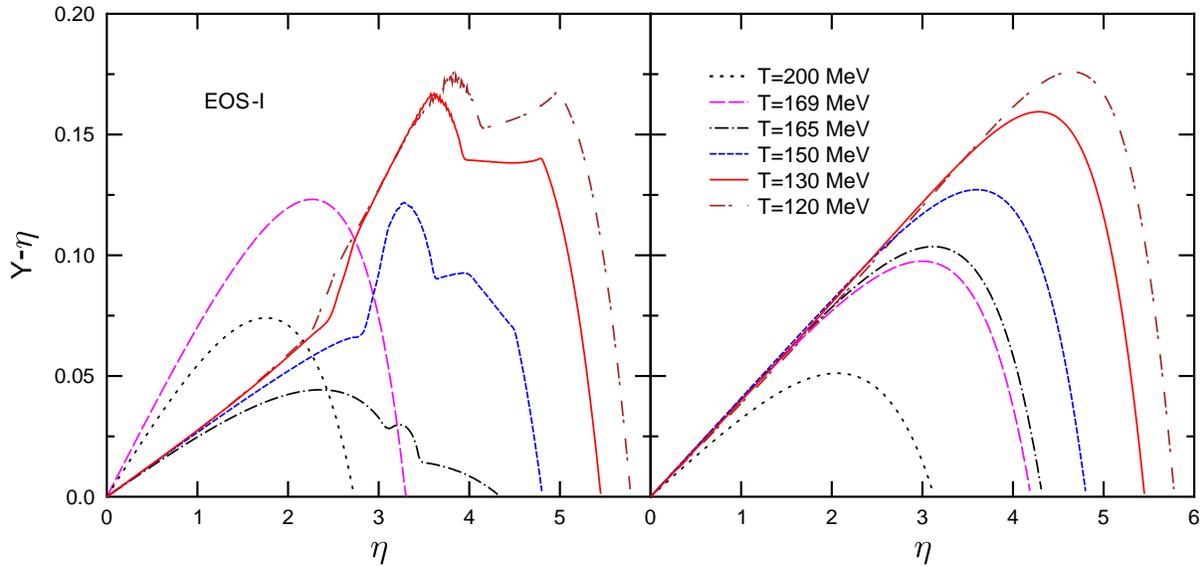}
\caption{
Same as Fig.~\ref{fig:eta-tauf}, but for profiles of collective rapidity
corresponding to fixed temperatures.
}
\label{fig:eta-yf}
\end{figure*}

Space--time evolution of matter as predicted by the present model is
illustrated in Figs.~\mbox{\ref{fig:eta-tem1}--\ref{fig:eta-yf}}.
In particular, Figs.~\ref{fig:eta-tem1}--\ref{fig:eta-y1} show profiles of the
temperature and the collective rapidity at different proper times
$\tau$\,. Here we consider the Gaussian--like initial conditions, with
parameters from the set A (see  Fig.~\ref{fig:sig-eta0} and
Table~\ref{tab3}). For comparison, the results  are presented for the
EOS--I and for the hadronic EOS with $c_s^2=0.15$. One can see
that in the case of the phase transition the model predicts appearance
of a a flat shoulder in $T\hsp(\eta)$ and local minima in $Y\hsp(\eta)$
which are clearly visible at $\tau\lesssim 10$\,fm/c. This is a
manifestation of the mixed phase which exists during the time interval
$\Delta\tau\sim 10$\,fm/c. According to Fig.~\ref{fig:eta-tem1}, the largest
volume of this phase in the $\eta$--space takes place at $\tau\sim 5$\,fm/c.
In the considered case the ''memory'' of the quark phase is practically
washed out at $\tau\gtrsim 30$\,fm/c. As one can see from
Fig.~\ref{fig:eta-y1}, at such late times deviation from the Bjorken scaling
($Y=\eta$) does not exceed~5\%.

Another important quantity which shows deviation from scaling
hydrodynamics is the entropy density $s$ multiplied by $\tau$. As
noted in footnote to Eqs.~(\ref{bjoe}), $s\tau$ is constant in the
Bjorken model. In Fig.~\ref{fig:eta-stau} we show profiles of $s\tau$
at different $\tau\geqslant\tau_0$\,, again for the parameter set A.
Two conclusions can be made here. First, these profiles are only weakly
sensitive to the presence of the phase transition. We have already
noted this when discussing Fig.~\ref{fig:eps-sden}. Second, one can
clearly see that $s\tau\neq\textrm{const}$ as a function of $\tau$\,.
In contrast to the Bjorken model, $s\tau$ drops by about 15\%
for $|\eta|<1$ at $\tau\gtrsim 20$\,fm/c. Due to this reason, we think
that the above--mentioned 2+1 dimensional models
\mbox{\cite{Kol99,Bas00,Per00,Kol01,Tea01}}, which assume Bjorken
scaling in the beam direction, are not very accurate even for the slice
around \mbox{$\eta=0$}.

To study sensitivity of flow to the EOS of the hadronic phase, we have
performed calculations for different sound velocities $c_H$ keeping other
parameters of the EOS the same as in
Figs.~\mbox{\ref{fig:eta-tem1}--\ref{fig:eta-stau}}. It is found that
at fixed total energy $E$\,, acceleration of the fluid is stronger for harder
EOS\hsp s characterized by larger~$c_H$\,. Figure~\ref{fig:eta-y2} shows profiles
of flow rapidity for the EOS\hsp s with $c_H^2=1/3$\footnote
{
As will be shown below, the observed pion and kaon yields can be
reproduced with the EOS--III after readjusting initial conditions.
As compared with ''standard'' calculations using $c_H^2=0.15$\,,
approximately the same quality of fit can be achieved by choosing
narrower profiles of initial energy density\,.
}.
Comparison with Fig.~\ref{fig:eta-y1} shows that the transition from
$c_H^2=0.15$ to $c_H^2=1/3$ results in about 3--fold increase of
relative flow rapidity $Y-\eta$\, (this quantity characterizes
deviations from the inertial expansion of matter). Comparison of
the left and right panels in Fig.~\ref{fig:eta-y2} reveals that the
sensitivity of flow to the phase transition is much stronger as
compared with the soft EOS considered above ($c_H^2=0.15$). The differences in
acceleration at late times of 30--50 fm/c are especially visible for
$|\eta|\lesssim 2$\,. Qualitative explanation of this effect is
obvious: fluid elements at such $\eta$ exhibit smaller acceleration in
the case of phase transition. This is a manifestation of the mixed phase
which is characterized by small gradients of pressure.

Figure~\ref{fig:eta-tauf} shows the fluid isotherms in the $\eta-\tau$
plane. The profiles of flow rapidity at hypersurfaces of constant
temperature are shown in Fig.~~\ref{fig:eta-yf}.  This information is
used to calculate particle momentum distributions by using formulae of
Sect.~\ref{sfreez}. According to Fig.~\ref{fig:eta-tauf}
(see the left panel), the initial stage of the evolution when matter is in the
quark--gluon phase, lasts only for a very short time, of about 5 fm/c.
The region of the mixed phase is crossed in less than 10 fm/c. This
clearly shows that the slowing down of expansion associated with the
''soft point'' of the EOS plays no role, when the initial state lies
much higher in the energy density than the phase transition region. In
this situation the system spends the longest time in the hadronic
phase. The freeze-out at $T_F=130$\,MeV requires an expansion time of
about 60 fm/c at $\eta=0$. This is certainly a very long time which is
seemingly in contradiction with some experimental findings. Indeed, the
interferometric measurements~\cite{Adl01} show much shorter times of
hadron emission, of the order of 10 fm/c.
As follows from our results, this discrepancy can not be
removed by considering other EOS or initial conditions. A considerable
reduction of the freeze--out times can be achieved by including the
effects of transverse expansion and chemical
nonequilibrium~\cite{Hir02}. However, this will not change essentially
the dynamics of the early stage ($\tau\lesssim 10$\,fm/c) when
expansion is predominantly one--dimensional. A more radical solution
would be an explosive decomposition of the quark--gluon plasma,
proposed in Ref.~\cite{Mis99}. This may happen at very early times,
right after crossing the critical temperature line, when the plasma
pressure becomes very small or negative. We shall consider this
possibility in a forthcoming publication.

\subsection{Rapidity spectra of secondary hadrons\label{rapsp}}

Below we show the results for rapidity distributions of $\pi$-- and
$K$--mesons as well as antiprotons produced in central Au+Au collisions
at $\sqrt{s_{\scriptscriptstyle NN}}=200$\,GeV. These results are compared
with data of the BRAHMS Collaboration~\cite{Bea04,Bea05} for most central
(\mbox{0\%--5\%}) collisions.
\begin{table}[htb!]
\caption{
Parameters of the initial states which give the best fits of the pion,
kaon and antiproton rapidity spectra observed in central Au+Au
collisions at $\sqrt{s_{\scriptscriptstyle NN}}=200$\,GeV. All sets
correspond to the EOS--I. $T_0$~denotes the maximum temperature at $\tau=\tau_0$\,.
$E_1$ and $E_3$ are total energies of produced particles within the
rapidity intervals $|y|<1$ and $|y|<3$, respectively.
}
\label{tab3}
\bigskip
\begin{ruledtabular}
\begin{tabular}{c|c|c|c|c|c|c|c}
~set~&~$\epsilon_0$\,(GeV/fm$^3$) &
$\sigma$ & $\eta_0$ &~$T_0$\,(MeV) &
~$E_1$\,(TeV) &~$E_3$\,(TeV)&~$E/S$\,(GeV)\\
\hline
A & 10 &~1.74 & 0    & 279 & 1.53 & 9.25 & 0.89\\
B & 9  &~1.50 &~0.62 & 271 & 1.54 & 9.59 & 0.86\\
C & 8  &~1.30 &~1.14 & 263 & 1.49 & 9.55 & 0.86
\\[1mm]
\end{tabular}
\end{ruledtabular}
\end{table}
\begin{figure*}[htb!]
\hspace*{-2cm}
\includegraphics[width=0.9\textwidth]{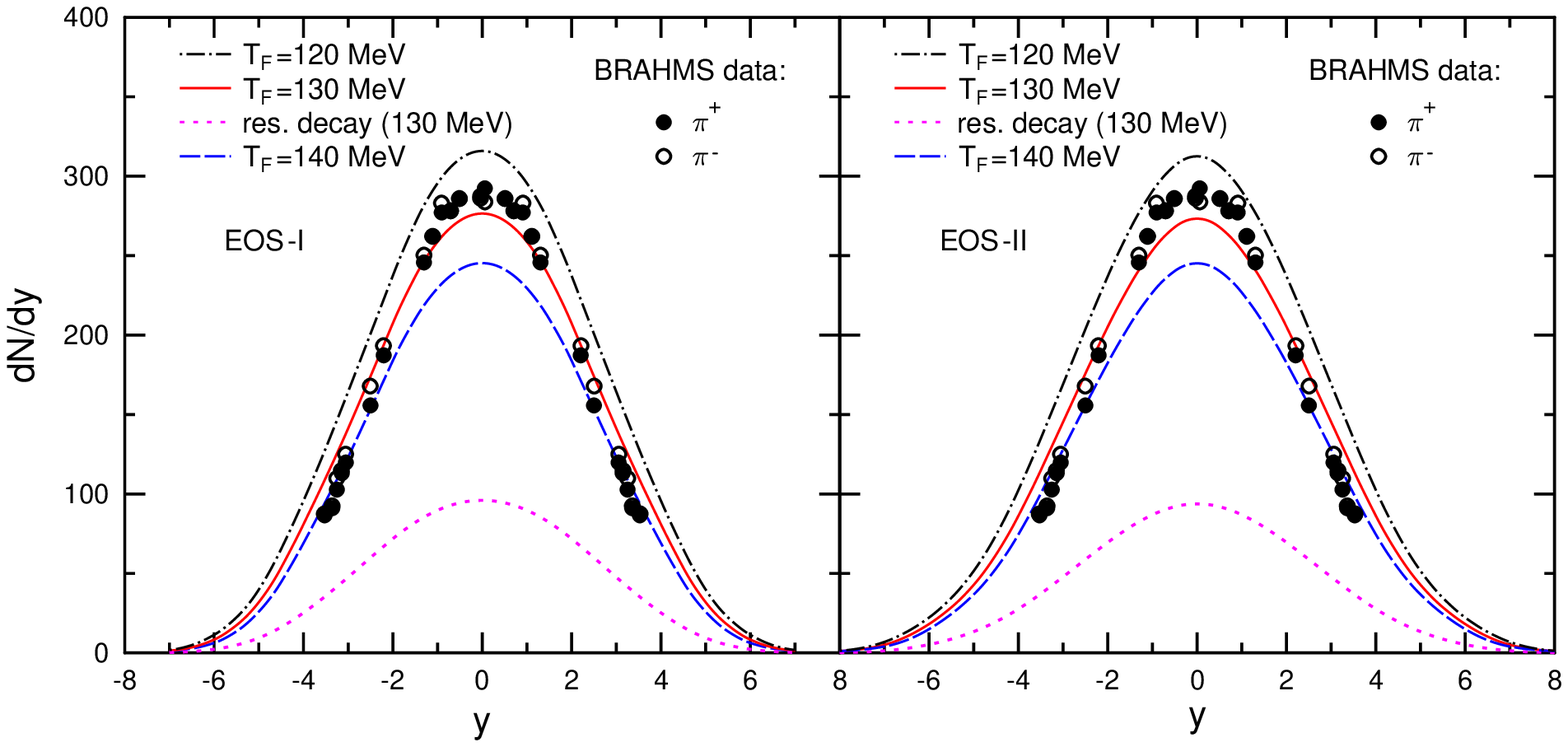}
\caption{
Rapidity distribution of $\pi^+$--mesons in central Au+Au collisions at
$\sqrt{s_{\scriptscriptstyle NN}}=200$\,GeV. Left panel shows
results of hydrodynamical
calculations for the EOS--I and initial conditions with parameters
\mbox{$\epsilon_0=10$\,GeV/fm$^3$, $\eta_0=0$, $\sigma=1.74$} (set A in
Table~\ref{tab3}). Right panel corresponds to the EOS--II and the
parameters \mbox{$\epsilon_0=5$\,GeV/fm$^3$, $\eta_0=0$,
$\sigma=2.02$}. Solid, dashed and dashed--dotted curves correspond to
different values of the freeze--out temperature $T_F$\,. The dotted
line shows contribution of resonance decays in the case $T_F=130$\,MeV.
Experimental data are taken from Ref.~\cite{Bea05}.
}
\label{fig:rap-pis1}
\end{figure*}

We have considered different profiles of the initial energy density,
ranging from the Gaussian--like ($\eta_0=0$) to the table--like
($\sigma=0$) initial states. We found that in the case of EOS--I it is not possible
to reproduce the BRAHMS data on the pion and kaon rapidity spectra in
Au+Au collisions by choosing either too small ($\epsilon_0\lesssim
5$\,GeV/fm$^3$) or too large ($\epsilon_0\gtrsim 15$\,GeV/fm$^3$)
initial energy densities. For such $\epsilon_0$ values  the pion and
kaon yields can not be reproduced with any $T_F$\,. According to our
calculations, the quality of fits is noticeably reduced for initial
energy density profiles with sharp edges, corresponding to
$\sigma<1$\,. As follows from the constraint~(\ref{teval}), such
profiles should have a very large $\epsilon_0$ or a significant plateau
$-\eta_0<\eta<\eta_0$. This would lead to more flat rapidity
distributions of pions and kaons as compared to the BRAHMS data (see
the next section).

A few parameter sets which lead to good fits with the EOS--I are listed
in Table~\ref{tab3}. In these calculations we choose various
$\epsilon_0$ and $\sigma$ and determine $\eta_0$ from the total energy
constraint~(\ref{teval}).
\begin{figure*}[htb!]
\hspace*{-15mm}
\includegraphics[width=0.9\textwidth]{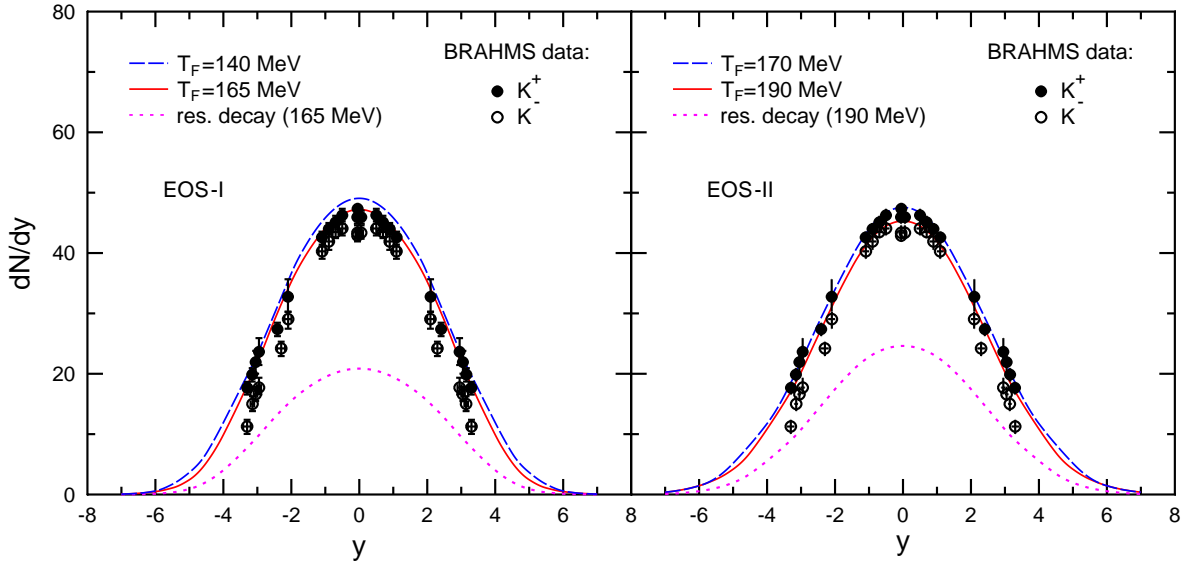}
\caption{
Same as Fig.~\ref{fig:rap-pis1}, but for $K^+$--meson rapidity
distributions.
}
\label{fig:rap-kas1}
\end{figure*}
It is interesting that the initial states A--C have approximately the
same total entropy $S\simeq 3\cdot 10^4$ (see Fig.~\ref{fig:sig-eta0}).
This is not surprising because the total entropy
should be proportional to the total
multiplicity of secondary mesons which is approximately same for all
good fits. As one can see from the last column of
Table~\ref{tab3}, the corresponding $E/S$--ratios fall into a narrow
interval $0.86-0.89$\,GeV. This observation is similar to the result of
Ref.~\cite{Cle98} that the observed ratio of the rest--frame energy to
the multiplicity of produced hadrons is constant as a function of the
bombarding energy.

To check the sensitivity to the parameters of the phase transition, we
have also calculated the pion and kaon rapidity distributions for the
EOS--II.  It is found that with the same initial energy profiles as for
the EOS--I, it is not possible to reproduce the observed spectra at any
freeze--out temperature. In particular, the predicted kaon yield is
strongly overestimated\hsp\footnote
{
We have checked that at fixed $T_F$ and the same initial conditions the
freeze--out times predicted for the EOS--II are noticeably longer than
for the EOS--I.
}
at \mbox{$100\,\textrm{MeV}<T_F<T_H=190$\,MeV}. Nevertheless, the
BRAHMS data can be well reproduced with the EOS--II too, when taking
smaller initial energy densities as compared with the EOS--I. Fits of
approximately same quality are obtained for
$\epsilon_0\simeq 5$\,GeV/fm$^3$\,. As before, in choosing the initial
conditions we apply the constraint~(\ref{teval}) for the total energy
of produced particles.  Similarly to the case of the EOS--I, the data
are better reproduced for initial profiles with small
$\eta_0\lesssim 1$\,.

Figures~\ref{fig:rap-pis1}--\ref{fig:rap-kas1} show the model results
for pion and kaon rapidity distributions obtained for the EOS--I and
EOS--II. These results correspond to Gaussian initial profiles with
$\eta_0=0$\,. For both EOS\hsp s we choose the parameter $\epsilon_0$
to obtain the best fit of the BRAHMS data\footnote
{
We did not try to achieve a perfect fit of these data, bearing in mind
that their systematic errors are quite big, about 15\% in the rapidity
region $|y|>1.3$~\cite{Bea05}.
}.
Although the overall fits are very similar for both EOS\hsp s, the
rapidity spectra obtained with the EOS--II are slightly broader than
those with the EOS--I. In the same figures we demonstrate sensitivity
to the choice of the freeze--out temperature. The best fits of the
pion spectrum are achieved with \mbox{$T_F\simeq 130$~MeV} (see
Fig.~\ref{fig:rap-pis1}). On the other hand, the kaon spectrum can be
well reproduced only by assuming that kaons decouple at the very
beginning of the hadronic stage, i.e. at
\mbox{$T_F\simeq T_H=165\,(190)$\,MeV} for EOS--I (II).
The contribution of resonance decays turns out to be rather
significant, especially in the central rapidity region, where it
amounts to about 35\% (45\%) of the total pion (kaon) yield.

\begin{figure*}[htb!]
\centerline{\includegraphics[width=0.8\textwidth]{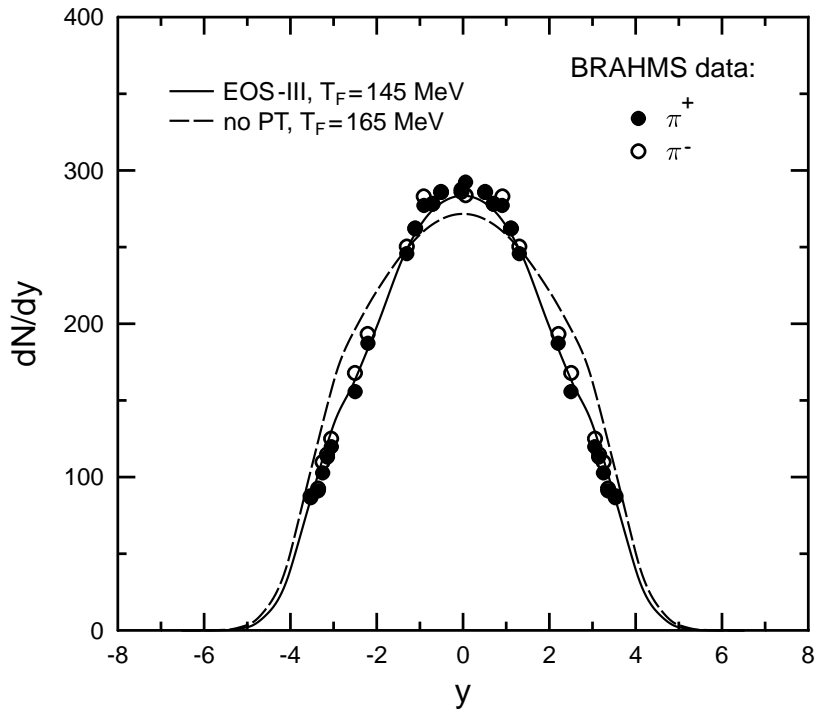}}
\vspace*{-5mm}
\caption{
Rapidity distribution of $\pi^+$--mesons in central Au+Au collisions at
\mbox{$\sqrt{s_{\scriptscriptstyle NN}}=200$} GeV calculated
for initial conditions with parameters
\mbox{$\epsilon_0=50$\,GeV/fm$^3$,} $\eta_0=0$,
$\sigma=0.96$.  The solid line corresponds to the EOS--III. The dashed
line is calculated for the hadronic EOS $P=\epsilon/3$\,.
Experimental data are taken from Ref.~\cite{Bea05}.
}
\label{fig:rap-pis3}
\end{figure*}
According to Fig.~\ref{fig:rap-pis1}, larger yields of secondary pions
are predicted for smaller freeze--out temperatures. A much weaker
sensitivity to $T_F$ is found for kaons (see Fig.~\ref{fig:rap-kas1}).
This difference can be explained by the large difference between the
pion and kaon masses.  Indeed, in the case of direct pions, a good
approximation at $T_F>100$\,MeV  is to replace the transverse mass
$m_T$ in~\re{ydis} by the pion transverse momentum~$p_T$. Neglecting
the term with $\tau^{\hsp\prime}_F$ in the r.h.s. of this equation, one
can show that the rapidity distribution of pions at $y=0$ is
proportional to
$\xi=\tau_F(\eta)\hsp\cosh{\eta}\cdot T_F^{\hsp
3}/\cosh^3{Y_F(\eta)}$ integrated over all $\eta$\,. For a rough
estimate, one can use the Bjorken relations~\cite{Bjo83}
$\mbox{$Y_F=\eta$},\, s_F\tau_F=s_0\tau_0$\,, where $s_F$ is the
entropy density at $T=T_F$\,. Using~\re{heos} one gets
\mbox{$\tau_F\propto s_F^{-1}\propto T_F^{-1/c_H^2}$} and therefore,
$\xi\propto T_F^{3-1/c_H^2}$. This shows that for $c_H^2<1/3$ the
pion yield grows with decreasing $T_F$. Qualitatively, one can say that
at low enough $c_H$  the increase of the spatial volume at freeze--out
compensates for the decrease of the pion occupation numbers at smaller
$T_F$. This effect is somewhat reduced because of a decreasing
resonance contribution at smaller temperatures. It is obvious that for
kaons this effect should be much weaker due to the presence of the
activation exponent $\exp{(-m_K/T_F)}$\,. In fact, a numerical
calculation for the same EOS and initial state shows that the kaon
yield changes nonmonotonically: first it slightly increases when
temperature goes down but then it starts to decrease at lower~$T_F$\,.

\begin{figure*}[htb!]
\vspace*{2mm}
\centerline{\includegraphics[width=0.8\textwidth]{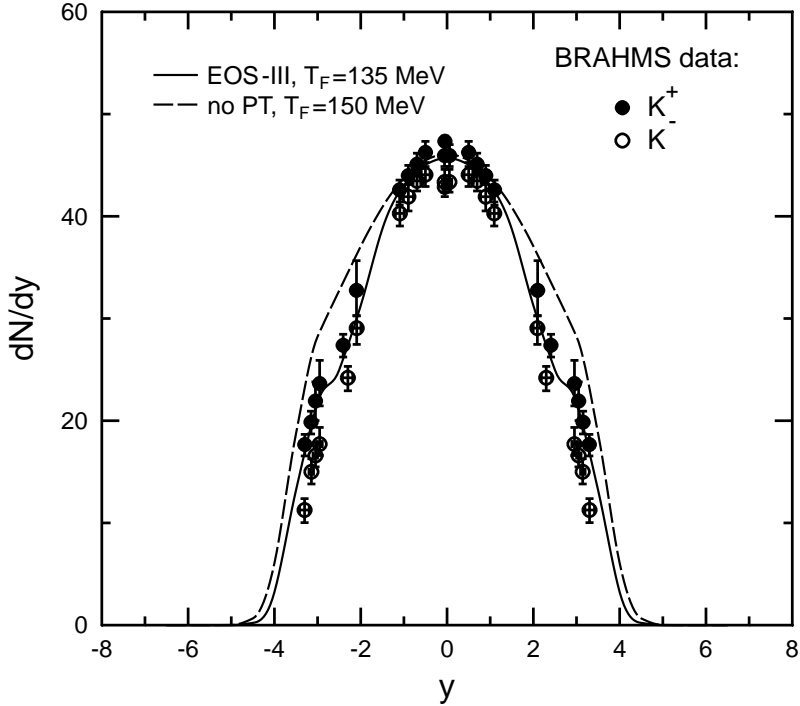}}
\caption{
Same as Fig.~\ref{fig:rap-pis3}, but for $K^+$--meson rapidity
distributions.
}
\label{fig:rap-kas3}
\end{figure*}
To investigate sensitivity of particle spectra to the presence of the
phase transition, we have performed calculations with purely hadronic
EOS\hsp s. In this case we use the same initial conditions as before,
but apply \re{heos} for all stages of the reaction, including the high
density phase. Our analysis shows that for soft hadronic EOS with
$c_H^2\lesssim 0.2$ it is possible to reproduce the observed pion and
kaon data with approximately the same fit quality as in the
calculations with the quark--gluon phase transition. Furthermore, the
corresponding freeze--out temperatures do not change significantly.
However, we could not achieve satisfactory fits for the ''hard''
hadronic EOS with \mbox{$c_H^2\geqslant 1/3$}.
Figures~\mbox{\ref{fig:rap-pis3}--\ref{fig:rap-kas3}} show particle
spectra calculated for the same EOS\hsp s and initial conditions as in
Fig.~\ref{fig:eta-y2}. One can see that the hadronic EOS with
$c_H^2=1/3$ leads to too wide rapidity distributions for both pions and
kaons. The reason is that the higher pressure in the hadronic EOS gives
a stronger push to the matter in forward and backward directions. From
these findings we conclude that a certain degree of softening of the
EOS is required to  reproduce the pion and kaon rapidity distributions.
Our analysis shows that reasonable fits of data for EOS\hsp s with
$c_H\gtrsim 0.2$\, (with and without the phase transition)
can be achieved only when freeze--out temperatures for kaons are
smaller than for pions. We find this situation unphysical, because
kaons have smaller rescattering cross sections and, therefore,
there is no reason for delayed kaon emission.
\begin{figure*}[htb!]
\hspace*{-1cm}
\includegraphics[width=0.9\textwidth]{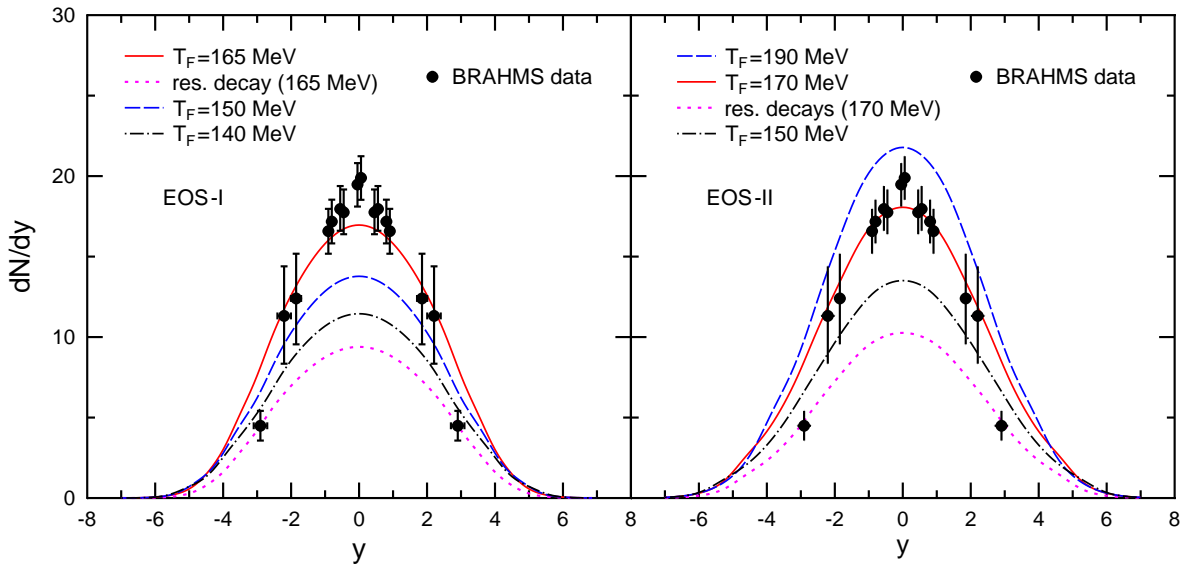}
\caption{
Same as Fig.~\ref{fig:rap-pis1}, but for antiproton rapidity
distributions. All results are obtained assuming $\mu_F=0$.
Experimental data are taken from Ref.~\cite{Bea04}.
}
\label{fig:rap-pbs}
\end{figure*}

It turned out that using the parameter sets from Table~\ref{tab3}, one can also
reproduce reasonably well the antiproton rapidity spectra measured by
the BRAHMS Collaboration~\cite{Bea04}.  Figure~\ref{fig:rap-pbs} shows
the antiproton rapidity distributions, calculated for the parameter
set A. In this case we explicitly take into account the contribution of the
\mbox{$\ov{\Delta}(1232)\to\pi\ov{p}$} decays, ignoring the width of
$\Delta$--isobar. Contributions of higher antibaryon resonances are
taken into account in a similar way as for pions and kaons. The
resonance contribution reaches about 55\% at $T_F=165$\,MeV. One should
consider these results as an upper bound for the antiproton yield. A
more realistic model should include the effect of nonzero baryon
chemical potentials which will certainly reduce the antibaryon yield.
The thermal model analysis of RHIC data, performed in
Refs.~\cite{And04,And06}, gives rather low values for the baryon
chemical potentials, $\mu_F\sim 30$\,MeV, at midrapidity. This will
suppress the antiproton yield by a factor~$\sim\exp{(-\mu_F/T_F)}\sim
0.8$\,.

\subsection{Rapidity distribution of total energy}

\begin{figure*}[htb!]
\centerline{\includegraphics[width=0.8\textwidth]{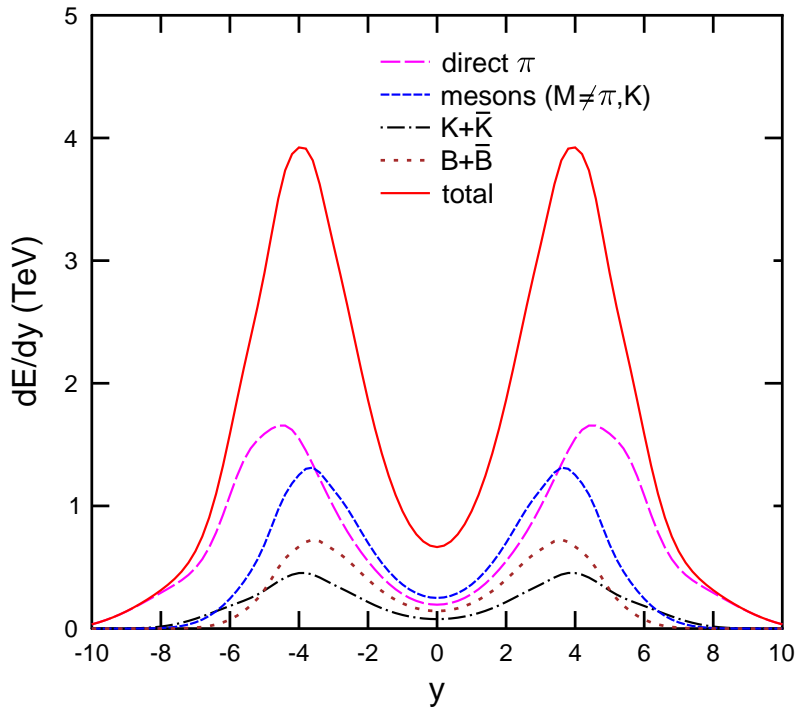}}
\caption{
Rapidity distribution of the total energy of secondary hadrons in
central Au+Au collisions at $\sqrt{s_{\scriptscriptstyle NN}}=200$\,GeV.
Shown are results for the EOS--I and parameter set A.
}
\label{fig:rap-dedy}
\end{figure*}
In order to check the energy balance in the considered reaction, we
have calculated additionally the rapidity distribution of the total
energy of secondary particles, $dE/dy$\,. In the zero--width
approximation this quantity can be written as
\bel{detdy}
\frac{dE}{dy}=\cosh{y}\sum\limits_i\int d^{\hsp 2}p_T \frac{d^3 N_i}
{d^{\hsp 2}p_T dy}\sqrt{m_i^2+p_T^{\,2}}\,,
\ee
where the sum goes over all hadronic species.
In calculating $dE/dy$ we take into account not only direct pions
($i=\pi$) and kaons ($i=K,\ov{K}$), but also 14 lowest mass mesons with
widths $\Gamma\lesssim 150$\,MeV~\cite{PDG04}:
\bel{mres}
M=\eta,\rho,\omega,\eta^{\,\prime},f_0,a_0,\phi,b_1,f_1,a_2,\eta\hsp (1295),
K^*,K_1,K^*_2\,.
\ee
We also include 12 strange and nonstrange baryon--antibaryon ($B\ov{B}$)
species:
\bel{bar}
B=N,N(1520),N(1535),\Delta,\Delta(1520),\Delta(1620),
\Lambda,\Lambda(1405),\Lambda(1520),\Sigma,\Sigma(1385),\Xi\,.
\ee

In calculating contributions of different species to $dE/dy$ in the
r.h.s. of~\re{detdy} we use the freeze--out temperatures obtained by
fitting the pion ($T_F^{\pi}$), kaon ($T_F^K$) and
antiproton~($T_F^{\ov{p}}$) rapidity distributions. We choose these
temperatures to find, respectively, the contributions of free pions~($i=\pi$),
heavy mesons ($i=M$\,, where $M\neq\pi$) and baryon--antibaryon
pairs ($i=B,\ov{B}$). In particular, the results shown in
Fig.~\ref{fig:rap-dedy} correspond to $T_F^\pi=130$\,MeV,
$T_F^K=T_F^{\ov{p}}=165$\,MeV. By integrating $dE/dy$, we have
determined $E_1$ and $E_3$, the total energies of secondaries within
the rapidity intervals $|y|<1$ and $|y|<3$, respectively. The BRAHMS
Collaboration estimated $E_{1,3}$ from the rapidity distributions of
charged pions, kaons, protons and antiprotons in most central Au+Au
collisions at $\sqrt{s_{\scriptscriptstyle NN}}=200$\,GeV. The values
$E_1\simeq 1.5\,\textrm{TeV},\hspace*{2ex}\mbox{$E_3\simeq 9\,\textrm{TeV}$}$
have been reported in Ref.~\cite{Ars05}.  From Table~\ref{tab3} one can see
that these values are well reproduced by the model.

Based on the above analysis we conclude that within the hydrodynamical
model the BRAHMS data can be well described with the EOS-I and EOS--II
and the parameters of the initial state ($\tau_0=1$\,fm/c)
$\sigma\simeq 1.5-2,~~\eta_0\lesssim 1$\,. The maximal initial energy
density, $\epsilon_0$, is sensitive to the critical temperature of the
phase transition. For the EOS--I ($T_c\simeq 167$\,MeV) we get the
estimate $\epsilon_0\simeq 9\pm 1$\,\textrm{GeV/fm}$^3$ while for the
EOS--II ($T_c\simeq 192$\,MeV) the required values of $\epsilon_0$ are
lower by about a factor of two.

\subsection{Comparison of Landau--like and Bjorken--like initial conditions
\label{blic}}

As discussed above, the initial profiles which give the best fits of
particle rapidity distributions are intermediate between the Landau and
Bjorken limits. Nevertheless it is interesting to compare the model predictions for
the Landau--like and Bjorken--like initial states. In fact, the Landau initial
conditions correspond to points situated near the origin of the
$\sigma-\eta_0$ plane in Fig.~\ref{fig:sig-eta0}. On the other hand, the
Bjorken--like states correspond to table--like states with small $\sigma$ and
large $\eta_0$ i.e. to points near the vertical axis in Fig.~\ref{fig:sig-eta0}\,,
away from the origin.

Let us first consider the limiting case of the Landau model.
\begin{figure*}[htb!]
\hspace*{-15mm}\centerline{\includegraphics[width=0.9\textwidth]{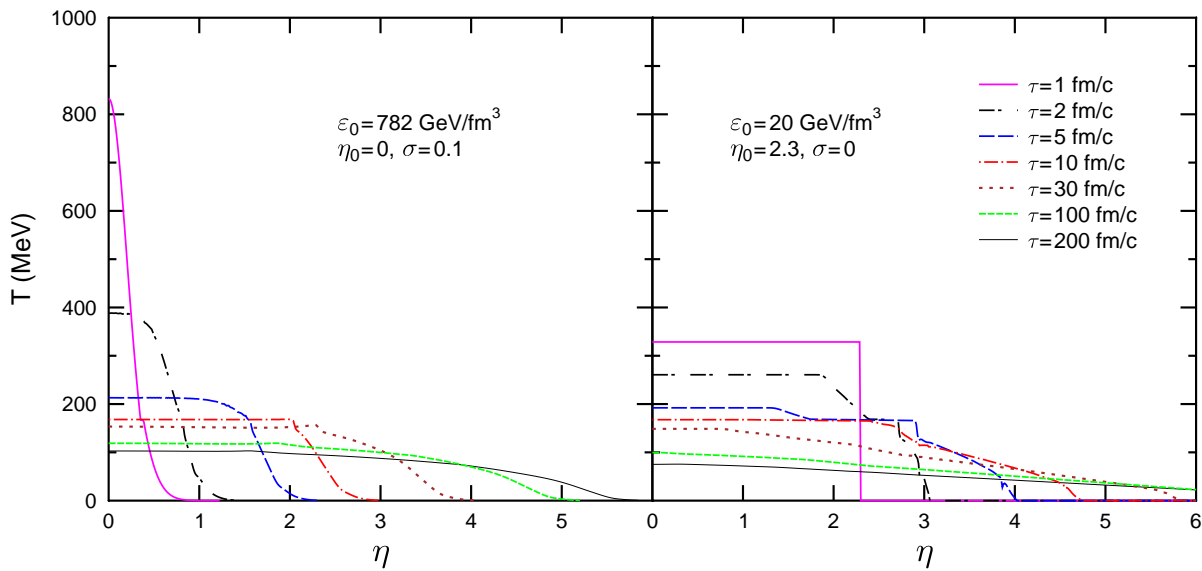}}
\caption{
Temperature profiles at different proper times $\tau$ calculated for
the EOS--III. Left~(right) panel corresponds to the Gaussian
(table-like) initial conditions with parameters given in the figures.
}
\label{fig:eta-teml}
\end{figure*}
In its simplest version, for the EOS\hsp s with $c_s={\rm const}$, this
model predicts a Gaussian rapidity distribution of secondary
pions. Originally, the Landau model was formulated for proton--proton collisions.
Below we assume that pion rapidity distributions in
$pp$--reactions and central collisions of equal nuclei have the same shapes
at the same c.m. energy $\sqrt{s_{\scriptscriptstyle NN}}$. Then the following
approximate formulae can be derived~\cite{Shu72}
\bel{lrdi}
\frac{dN_\pi}{dy}=C_\pi\frac{s_{\scriptscriptstyle NN}^{\,\alpha}}
{\sigma_{\rm Lan}}
\exp{\left[-\frac{y^2}{~~2\hsp\sigma_{\rm Lan}^2}\right]},
\ee
where the normalization constant $C_\pi$ does not depend on the
bombarding energy and
\bel{lpar}
\alpha=\frac{1-c_s^2}{2\hsp (1+c_s^2)}\,,~~\sigma_{\rm Lan}^2\simeq\frac{8}
{3}\frac{\ds c_s^2}{\ds 1-c_s^4}\ln{\frac{\sqrt{s_{\scriptscriptstyle NN}}}
{2\hsp m_N}}\,.
\ee
For $c_s^2=1/3$ and $\sqrt{s_{\scriptscriptstyle NN}}=200$\,GeV one obtains
$\alpha=1/4$ and $\sigma_{\rm Lan}\simeq 2.16$\,, the values often quoted in the
literature~\cite{Bea05,Ble05}. On the other hand, for $c_s^2=0.15$\,,
which is preferable from the viewpoint of our previous analysis (see
Sect.~\ref{rapsp}), the predicted width is only~1.38 i.e. significantly smaller
than observed by the BRAHMS Collaboration. Note, that the analytical
formulae~(\ref{lrdi})--(\ref{lpar}) are obtained for the case of
constant sound velocity and can not be applied for EOSs with
a first--order pase transition. In this connection we think that
attempts~\cite{Ble05} to extract a unique sound velocity from pion spectra
observed at a specific bombarding energy are unjustified.
\begin{figure*}[htb!]
\hspace*{-15mm}\centerline{\includegraphics[width=0.9\textwidth]{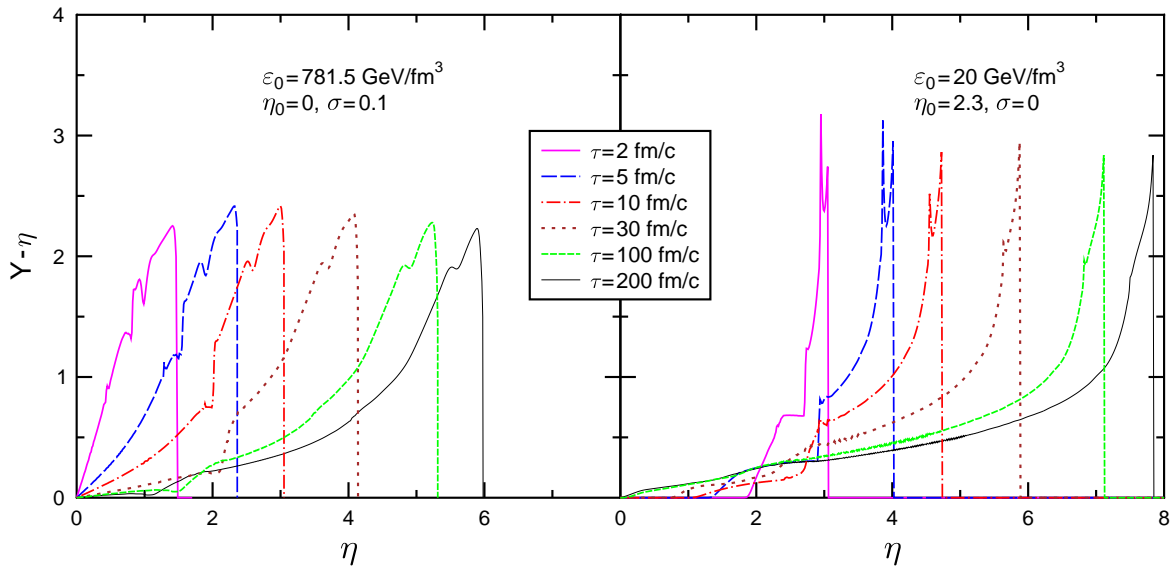}}
\caption{
Same as Fig.~\ref{fig:eta-teml}, but for collective rapidity profiles.
}
\label{fig:eta-yl}
\end{figure*}

Within the Landau model one assumes that a thin disk--like fireball is
created at some time $t=t_0$ after beginning of the nuclear
collision\hsp\footnote
{
In fact, in the Landau model formulated in cartesian coordinates, one
can choose $t_0$ arbitrary. However, in order to map the Landau--like
initial state on the $\tau-\eta$ coordinates, the parameter $\tau_0$ should not
be too small (see below).
}.
It is postulated that this fireball is initially at rest ($Y=0$)
in the c.m. frame. It has the transverse cross section
\mbox{$S_\perp=\pi R^2$} and the longitudinal size $\Delta z\sim
2R/\gamma$\,. Here $R$ is the geometrical radius of colliding nuclei,
$\gamma\sim\sqrt{s_{\scriptscriptstyle NN}}/2m_N$ is the c.m. Lorentz--factor.
The total energy, $E$\,, is normally assumed to be $E=\kappa\hsp (\gamma-1)\hsp
N_{\rm part}m_N$ where $N_{\rm part}$ is the number of participating
nucleons and \mbox{$\kappa\sim 0.5-0.7$} is the so--called inelasticity
coefficient.  At the c.m. energy \mbox{$\sqrt{s_{\scriptscriptstyle NN}}=200$\,GeV}
the Lorentz--factor $\gamma\sim 100$ and, therefore, $\Delta z\lesssim 0.1$\,fm.
If $t_0$ is not too small, the initial fireball occupies a region
of small space--time rapidities, \mbox{$|\eta|<\Delta\eta\simeq\Delta z/t_0$},
and one can approximately apply the parametrization of initial state (\ref{incd})
with $\eta_0=0$ and \mbox{$\sigma\sim 2R/(\gamma t_0)$}. Note, that at
small~$|\eta|$ the difference between the initial flow profiles
$Y=\eta$ and $Y=0$ is negligible. Using further~\re{tener2}, one can
estimate the maximal energy density of the fireball as $\epsilon_0\sim
E\,(\sqrt{2\pi^3}R^2\sigma t_0)^{-1}$\,.  On the basis of these
arguments we conclude that the Landau model can be approximately
simulated within our approach by taking the Gaussian initial profile with small
$\sigma\propto\gamma^{-1}$ and large $\epsilon_0\propto\gamma^2$\,.

\begin{figure*}[htb!]
\hspace*{-15mm}\includegraphics[width=0.9\textwidth]{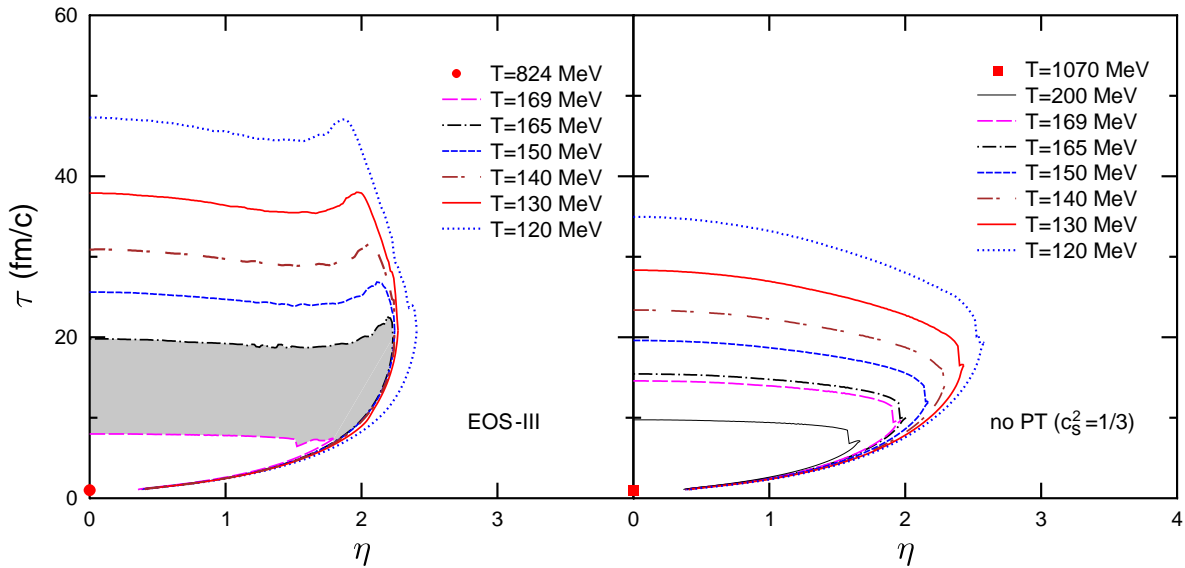}
\caption{
Isotherms in the $\eta-\tau$ plane calculated for the initial
parameters \mbox{$\epsilon_0=782$\,\,GeV/fm$^3$}, \mbox{$\eta_0=0$\,,
$\sigma=0.1$}.
Left panel corresponds to the EOS--III. Right panel shows results
for the hadronic EOS $P=\epsilon/3$. Shaded region indicates the mixed
phase.  Dots near the origin correspond to the maximal initial temperature.
}
\label{fig:eta-taufl}
\end{figure*}
For illustration, in this section we present the numerical results for
the parameters of the initial state $\eta_0=0,\,\sigma=0.1$ and
$t_0\simeq\tau_0=1$\,fm/c. Again, the total energy is fixed at the
value~(\ref{teval}). This gives the estimate
$\epsilon_0\simeq 782$\,GeV/fm$^3$\,.
Figures~\ref{fig:eta-teml}--\ref{fig:eta-yl} show
profiles of the temperature and flow rapidity at different proper times
$\tau\geqslant\tau_0$\,.  For comparison with the Bjorken--like
scenario, we also present results for a ''broad'', table--like initial
state with $\sigma=0,\,\eta_0=2.3,\,\epsilon_0=20$\,GeV/fm$^3$ (this
state has the same total energy and entropy).  The difference between
the fluid dynamics predicted for the Landau--like and Bjorken--like initial
states is clearly visible at $\tau\lesssim 10$\,fm/c and $|\eta|\gtrsim
2$\,.  It is interesting that asymptotically, for $\tau\gtrsim
100$\,fm/c, the ''Landau'' profiles at $|\eta|\lesssim 2$ behave
similar to Bjorken's scaling solution, $Y\simeq\eta,\,
T\simeq\textrm{const}$\,. On the other hand, the two solutions
differ significantly at large $\eta$\,, where the Bjorken--like initial
state leas to a very strong acceleration of the fluid near its outer
edge.

Figure~\ref{fig:eta-taufl} shows contours of constant temperature for
the Landau--like initial state. These contours are calculated for two EOS\hsp s,
with and without the phase transition.
Contrary to the case of a broader initial energy density profile presented in
Fig.~\ref{fig:eta-tauf}, now the contours have a ''pocket--like'' structure
with back--bending portions of isotherms. One can see a moderate sensitivity
of isotherms to the presence of a phase transition. As discussed in
Sect.~ \ref{devm}, this sensitivity becomes even weaker for softer EOS\hsp s.

\begin{figure*}[htb!]
\centerline{\includegraphics[width=0.8\textwidth]{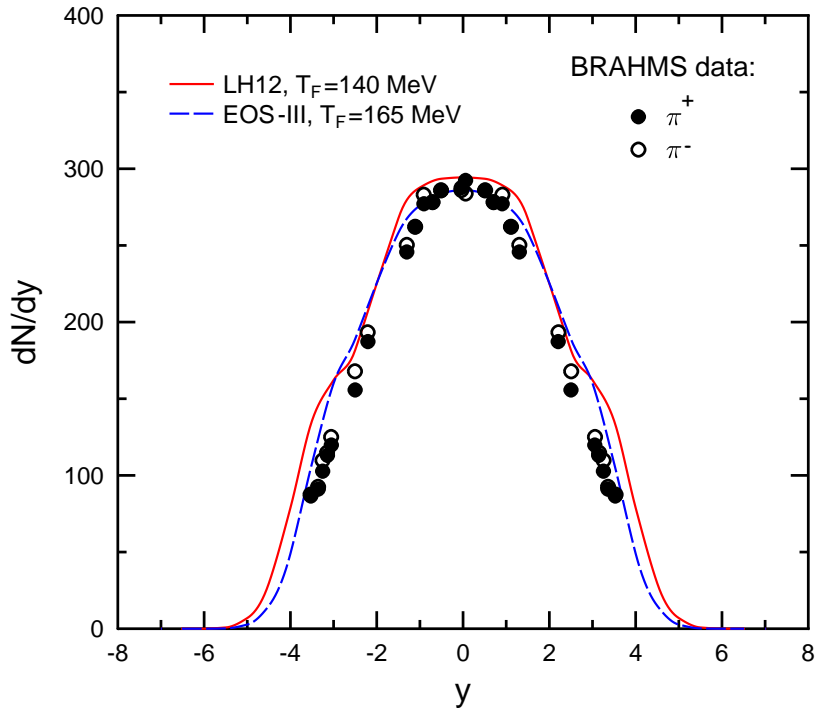}}
\caption{
Rapidity distribution of $\pi^+$--mesons in central Au+Au collisions at
\mbox{$\sqrt{s_{\scriptscriptstyle NN}}=200$\,GeV}. Shown are results of hydrodynamical
calculations for initial state with parameters \mbox{$\epsilon_0=782$\,GeV/fm$^3$},
$\eta_0=0$, $\sigma=0.1$\,. Experimental data are taken from Ref.~\cite{Bea05}.
}
\label{fig:rap-pisl}
\end{figure*}

\begin{figure*}[htb!]
\centerline{\includegraphics[width=0.8\textwidth]{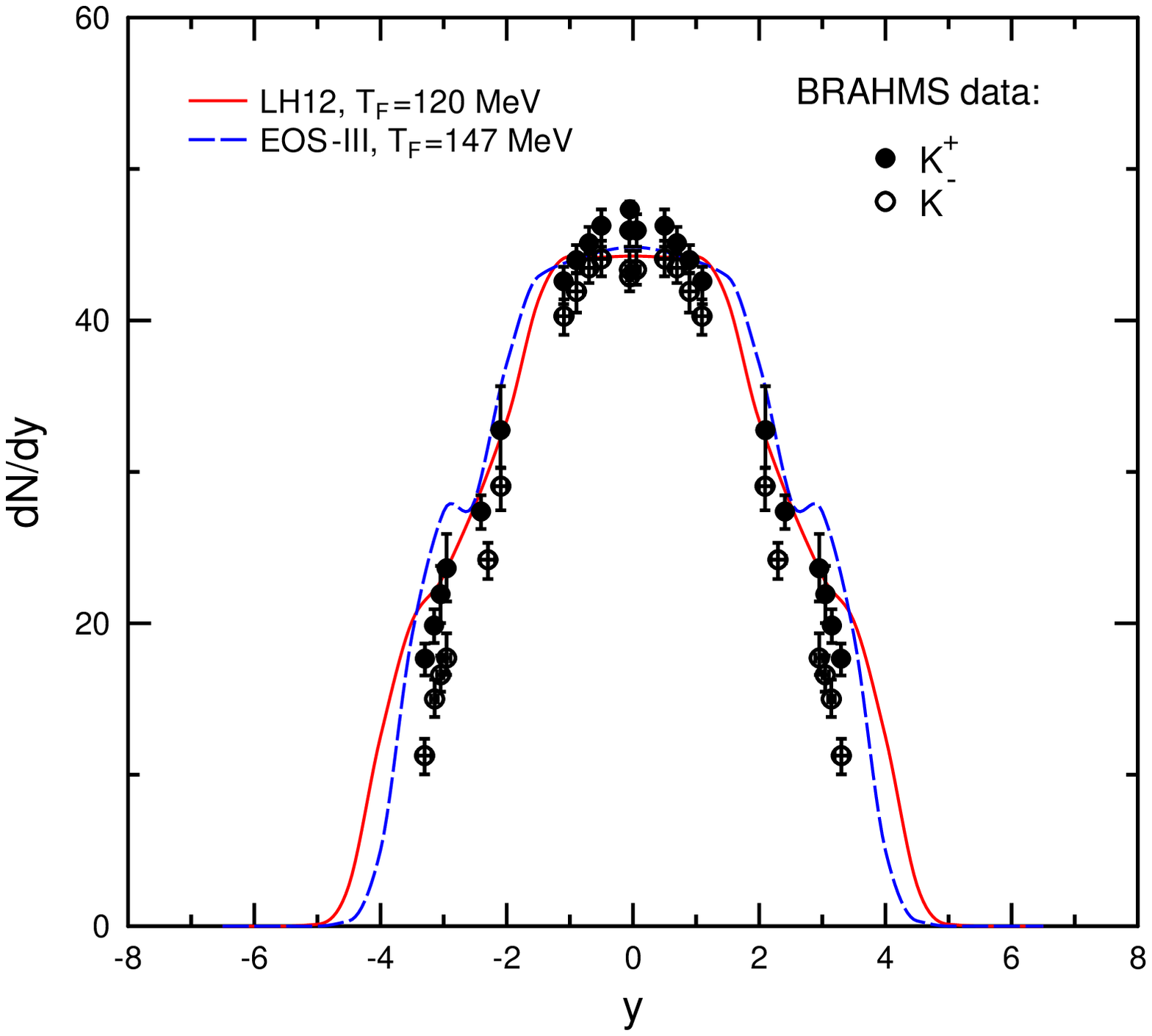}}
\caption{
Same as Fig.~\ref{fig:rap-pisl}, but for $K^+$--meson
rapidity distributions.
}
\label{fig:rap-kasl}
\end{figure*}

Now we present the rapidity distributions of pions and kaons
calculated with the Landau--like initial conditions for several EOS\hsp s.
Our analysis shows that it is not possible to reproduce the pion spectra
with EOS--I and EOS--II. The corresponding pion yields are too high at
any $T_F<T_H$\,. Both the pion and kaon yields become acceptable only
for EOS\hsp s with $c_H^2\gtrsim 0.2$\,. This is demonstrated in
Figs.~\ref{fig:rap-pisl}--\ref{fig:rap-kasl} which
show the pion and kaon spectra for two EOS\hsp s with $c_H^2=1/5$ and 1/3.
The first EOS corresponds to the EOS--I, but with $c_H^2=0.2$
(this is in fact the EOS LH12 from Ref.~\cite{Tea01}). In Fig.~\ref{fig:rap-pisl}
one can see that the pion spectrum calculated with
this EOS is somewhat broader as compared with the experimental data.
On the other hand, Fig.~\ref{fig:rap-kasl} shows that for the
EOS--III ($c_H^2=1/3$) the predicted kaon spectrum is noticeably
broader than observed experimentally.  Therefore, we
conclude that the BRAHMS data disfavor the Landau--like initial
conditions at RHIC bombarding energies.

Now let us consider in more details the Bjorken--like
initial conditions corresponding to the table--like profiles
($\sigma=0$) of energy density\hsp\footnote
{
Some results for such initial states have been already presented
in Figs.~\ref{fig:eta-teml}--\ref{fig:eta-yl} (right panels).
}.
In this case the energy density at $\tau=\tau_0$ is parametrized as
$\epsilon_0\hsp\Theta(\eta_0-|\eta|)$\,.
At fixed $\eta_0$ the parameter~$\epsilon_0$ can be determined
from the total energy constraint~(\ref{teval}). The explicit
expression for $\eta_0$ as a function of $\epsilon_0$
is given by~\re{etas} with $\alpha=0,\,\beta=1$\,.
Note, that strictly speaking, the Bjorken limit corresponds to infinite total energy and
\mbox{$\eta_0\to\infty$}\,.
Our calculations show that deviations from the Bjorken scaling increase
with decreasing $\eta_0$\,.

\begin{figure*}[htb!]
\centerline{\includegraphics[width=0.8\textwidth]{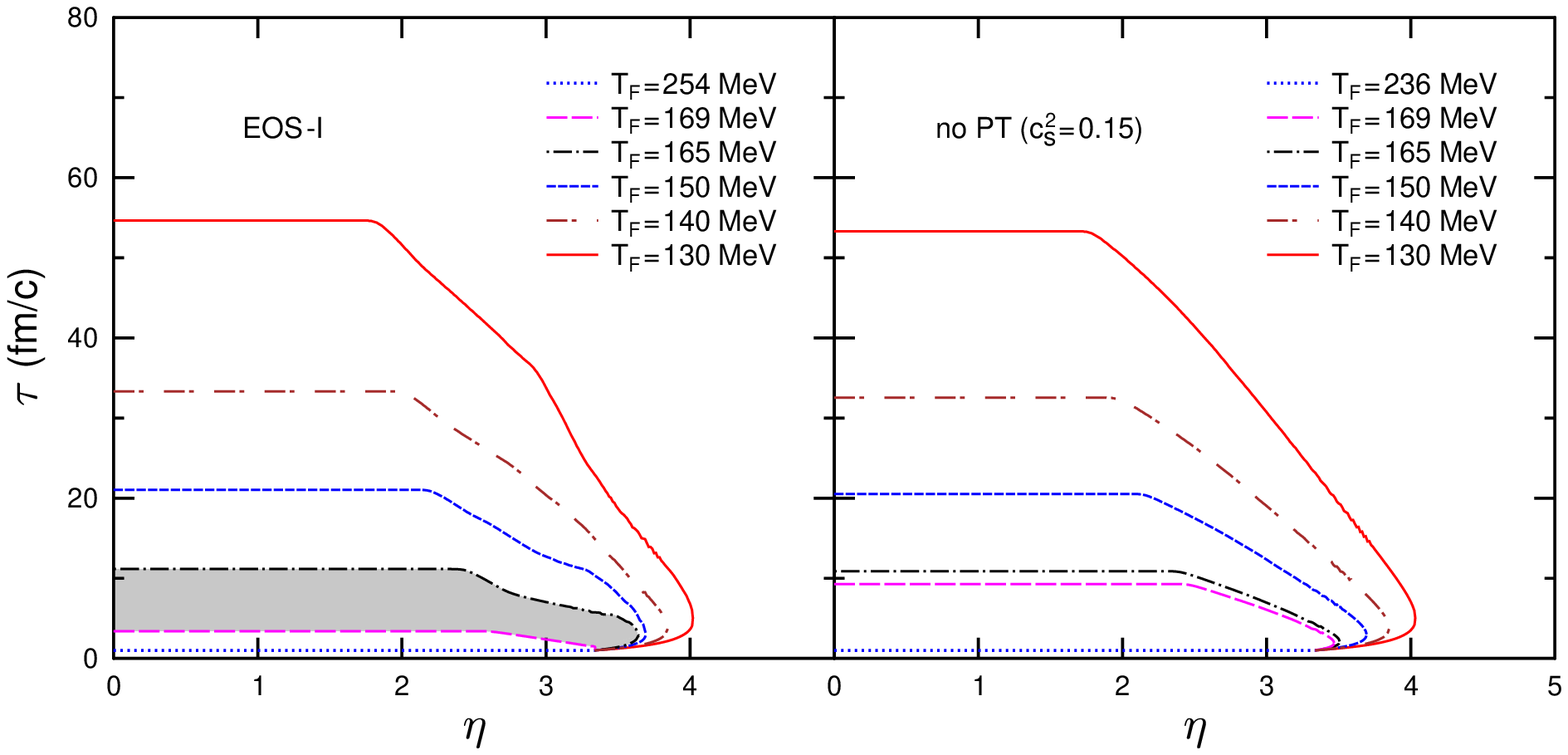}}
\caption{
Same as Fig.~\ref{fig:eta-tauf}, but for the initial profile
with parameters $\epsilon_0=7$\,GeV/fm$^3$\,, $\sigma=0$\,,
$\eta_0=3.34$\,.
}
\label{fig:eta-tauf1m}
\end{figure*}
Figure~\ref{fig:eta-tauf1m} shows isotherms for a table--like initial
state corresponding to the point D in Fig.~\ref{fig:sig-eta0}. Similarly to the
case of Landau--like initial conditions, these isotherms have
parts with inverse slope, but they are much smaller and appear only at large
$\eta$\,. The calculations show that these parts contribute only to the tails
of rapidity distributions.

\begin{figure*}[htb!]
\hspace*{-15mm}\centerline{\includegraphics[width=0.9\textwidth]{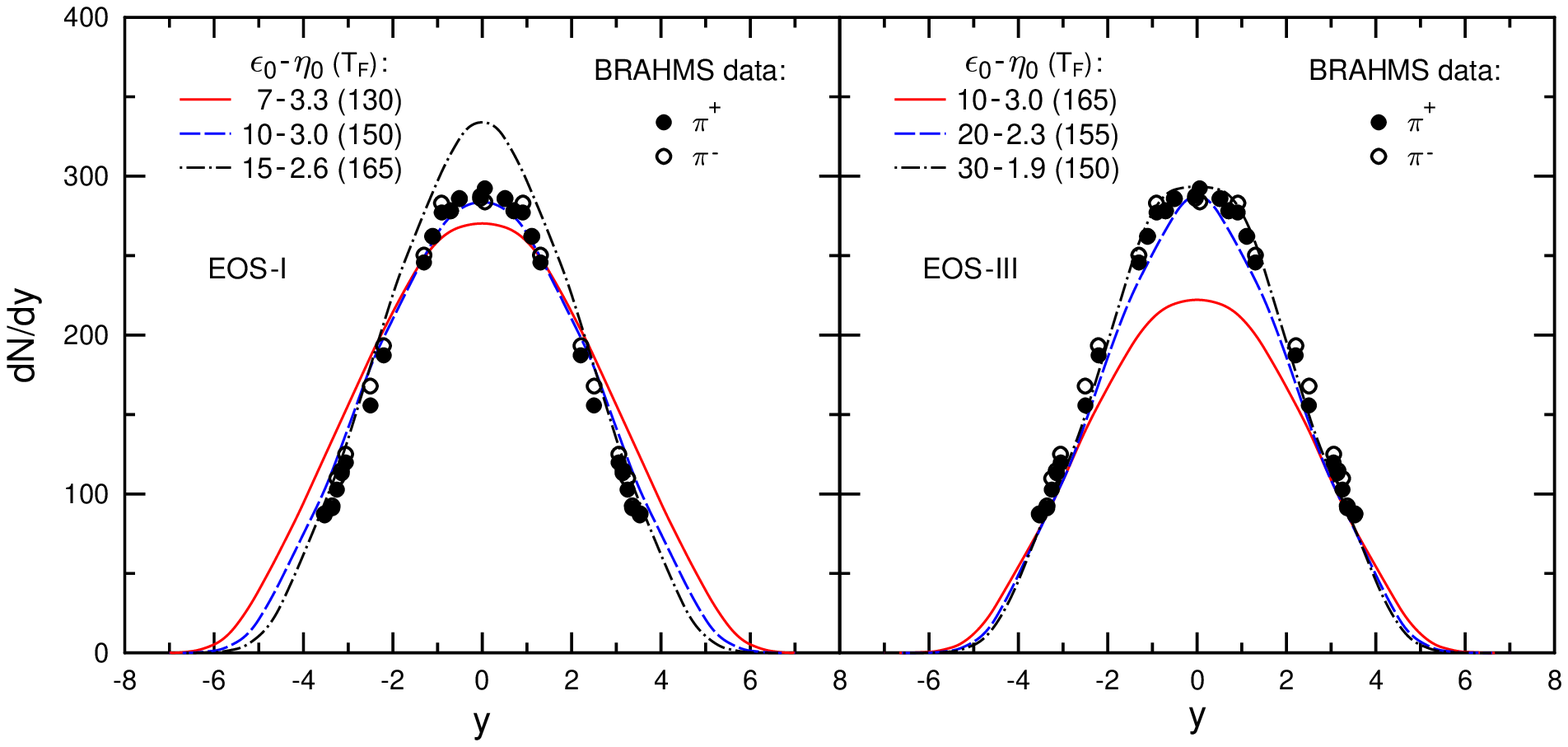}}
\caption{
Rapidity distribution of $\pi^+$--mesons in central Au+Au collisions at
$\sqrt{s_{\scriptscriptstyle NN}}=200$\,GeV. Solid, dashed and dashed--dotted
curves are calculated  for $\sigma=0$ and different
values of $\epsilon_0$~(in GeV/fm$^3$), $\eta_0$ and $T_F$ (in MeV).
The left (right) panel corresponds to the EOS--I~(III).
Experimental data are taken from Ref.~\cite{Bea05}.
}
\label{fig:rap-pis4}
\end{figure*}
\begin{figure*}[htb!]
\hspace*{-12mm}\centerline{\includegraphics[width=0.9\textwidth]{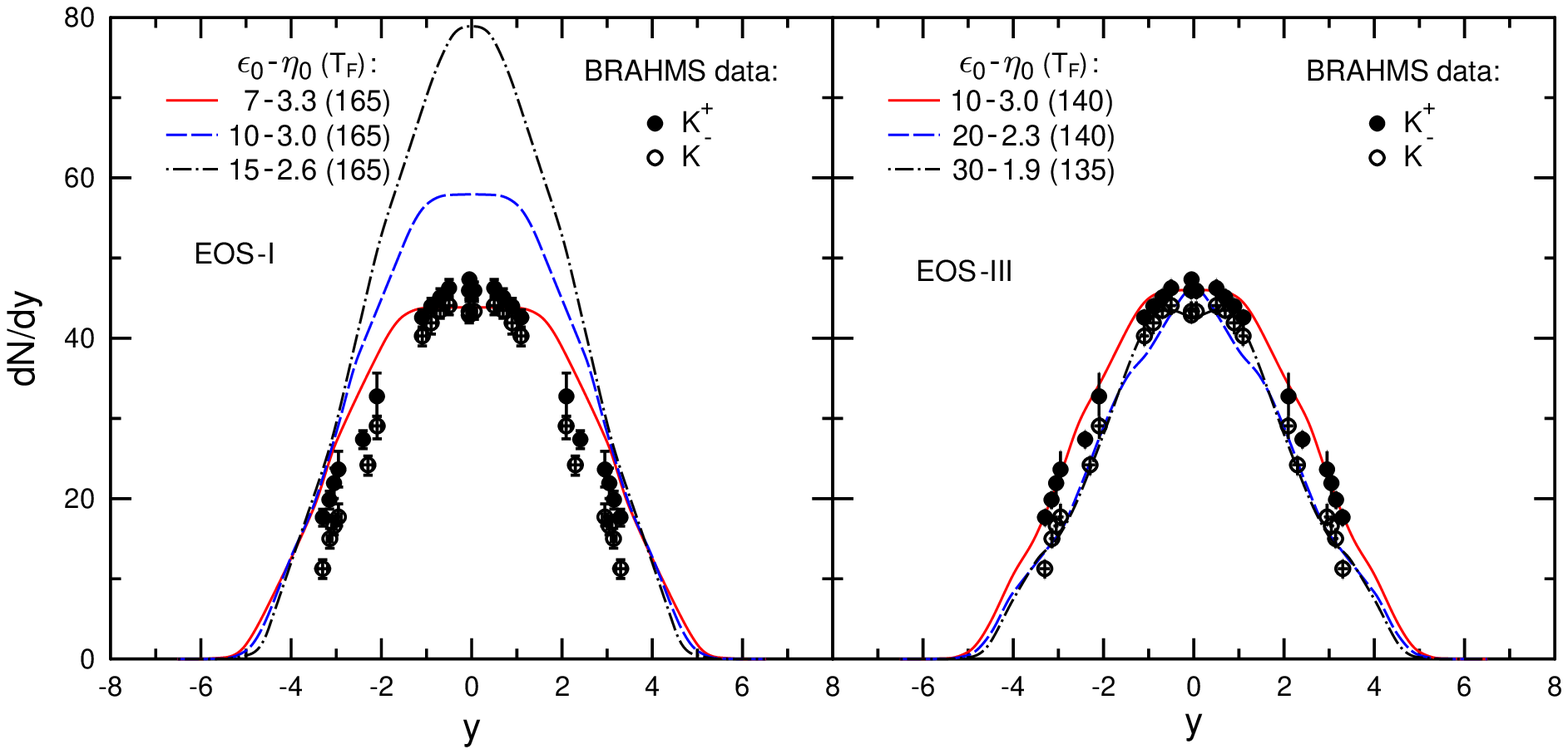}}
\caption{
Same as Fig.~\ref{fig:rap-pis4}, but for $K^+$--meson
rapidity distributions.
}
\label{fig:rap-kas4}
\end{figure*}
The pion and kaon rapidity distributions for Bjorken--like
initial states are given in Figs.~\ref{fig:rap-pis4}--~\ref{fig:rap-kas4}.
Figure~\ref{fig:rap-pis4} shows the pion rapidity distributions
calculated for several values of the parameter $\epsilon_0$ and for
two EOS\hsp s with the phase transition. In all cases
the freeze--out temperature is chosen by the best (when possible) fit
of BRAHMS data. In accordance with the
discussion in Sect.~\ref{rapsp}, the pion yield, calculated for
fixed initial state, decreases (increases) with raising $T_F$ for the
EOS--I~(III).

The results for kaon spectra are given in Fig.~\ref{fig:rap-kas4}.
One can see that at $\epsilon_0\gtrsim 10$\,GeV/fm$^3$
the produced kaon yield is overestimated for the EOS--I
at any freeze--out temperatures below $T_H=165$\,MeV
(we do not consider unrealistically small temperatures, $T_F\lesssim 50$\,MeV).
On the other hand, at lower $\epsilon_0$ both pion and kaon spectra become
too flat at midrapidity.
As seen from the right panels of Figs.~\ref{fig:rap-pis4}--\ref{fig:rap-kas4},
in the case of the EOS--III it is possible to reproduce simultaneously pion and
kaon spectra at $\epsilon_0\sim 20$\,GeV/fm$^3$\,. However, for such
$\epsilon_0$ the freeze--out temperatures for kaons are considerably lower
than for pions. This means that kaons are emitted later
than pions, which is apparently unrealistic.

On the basis of this analysis we conclude that both the Landau--like and
the Bjorken--like initial conditions lead to unsatisfactory results for
all considered EOS\hsp s.

\section{Summary and discussion}\label{scon}

In this paper we have generalized Bjorken's scaling hydrodynamics for
finite--size profiles of energy density in pseudorapidity space.  The
hydrodynamical equations were solved numerically in $\tau-\eta$
coordinates starting from the initial time $\tau_0=1$ fm/c until the
freeze--out stage. The sensitivity of the final particle distributions
to the initial conditions, the freeze--out temperature and the EOS has
been investigated. A comparison of $\pi, K, \ov{p}$\ rapidity spectra
with the BRAHMS data for central Au+Au collisions at
$\sqrt{s_{\scriptscriptstyle NN}}=200$\,GeV has been made. The best agreement with these data
is obtained for initial states with nearly Gaussian profiles of the
energy density. In choosing the initial conditions we impose the
constraint on the total energy of produced particles known from
experimental data~\cite{Bea04}. It is found that the maximum energy
density of the initial state, $\epsilon_0$, is sensitive to the
parameters of a possible deconfinement phase transition. The BRAHMS data are
well reproduced with $\epsilon_0$ of about 10 (5)~GeV/fm$^3$ for the
critical temperature $T_c\sim 165 \,(190)$\,MeV. The only
unsatisfactory aspect of these calculations is the prediction of very
long freeze--out times, $\sim 50$\,fm/c, for pions.

We would like to comment on several points. First, all calculations in
this paper are made for $\tau_0=1$\,fm/c. Of course, one can start the
hydrodynamical evolution from an earlier time, i.e. assuming smaller
$\tau_0$\,. In this case one should choose accordingly higher initial
energy densities. But $\tau_0$ cannot be taken too small, since at
very early times the energy is most likely stored in strong
chromofields~\cite{McL94}. The quark--gluon plasma is produced as a
result of the decay of these fields. Estimates show that the
characteristic decay times are in the range $0.3-1.0$~fm/c. At earlier
times the system will contain both the fields as well as produced
partons, and the evolution equations will be more complicated, see e.g.
Ref.~\cite{Mis02}.

It is obvious that the Cooper--Frye scenario of the freeze-out process,
applied in this paper is too simplified. This was demonstrated e.g. in
Ref.~\cite{Bra99}. One should also bear in mind that the freeze--out
temperatures obtained in our model will be modified by the effects of
transverse expansion and chemical nonequilibrium. Attempts to
construct a more realistic description of the freeze--out have been
performed in Refs.~\cite{Bas00, Tea01}, where the transport model was
applied to describe evolution of the hadronic phase. In this approach
the solution of fluid--dynamical equations is used to obtain initial
conditions for hadronic cascade calculations at later stages of matter
expansion. We are planning to study these problems in the future.

\appendix

\section{The fluid--dynamical equations in the conservative form}

In order to apply the standard flux--corrected transport
algorithm~\cite{Bor73} to solve numerically the equations of fluid
dynamics it is necessary to represent them in the conservative form.
The following set of equations can be written within the 1+1
dimensional model in the cartesian coordinates $t-z$
(for completeness we consider the general case of nonzero baryon charge)
\bel{conf}
\partial_t \bm{u}+\partial_z\hsp (\bm{u}v_z+\bm{f})+\bm{g}=0\,,
\ee
where $z$ and $v_z=U^z/U^0$ are the spacial coordinate and the
fluid 3--velocity. The multi--component vector $\bm{u}$ includes
the densities of baryon charge ($u_1=nU^0$), energy (\mbox{$u_2=T^{00}$}) and momentum
(\mbox{$u_3=T^{0z}$}). In the usual space--time representation
$\bm{g}$ vanishes and \mbox{$f_1=0$}, $f_2=Pv_z,\,f_3=P$\,.

To rewrite these equations in the light--cone variables, one can perform
the Lorentz--transformation of the baryon current $J^\mu=nU^\mu$ and the energy--momentum
tensor $T^{\mu\nu}$ to the frame moving with the rapidity $\eta=\rm{tanh}^{-1}(z/t)$
(this reference frame is marked below by tilde). In particular,
$\widetilde{J}^{\,0}=n\cosh{\widetilde{Y}}, \widetilde{J}^z=n\sinh{\widetilde{Y}}$ and
\begin{eqnarray}
\label{t00l}
&&\widetilde{T}^{00}=\epsilon\hsp\cosh^2{\widetilde{Y}}+P\sinh^2{\widetilde{Y}}\,,\\
\label{t0zl}
&&\widetilde{T}^{0z}=(\epsilon+P)\hsp\cosh{\widetilde{Y}}\sinh{\widetilde{Y}}\,,\\
\label{tzzl}
&&\widetilde{T}^{zz}=\epsilon\hsp\sinh^2{\widetilde{Y}}+P\cosh^2{\widetilde{Y}}\,,
\end{eqnarray}
where $\widetilde{Y}\equiv Y-\eta$\,.

From Eqs.~(\ref{bcce})--(\ref{emte}) of the main text we have
\begin{eqnarray}
\label{bccl}
&&\partial_\tau\widetilde{J}^{\,0}+
\partial_x\widetilde{J}^z+\widetilde{J}^{\,0}/\tau=0\,,\\
\label{ec1l}
&&\partial_\tau\tilde{T}^{00}+
\partial_x\widetilde{T}^{0z}+(\widetilde{T}^{00}+\widetilde{T}^{zz})/\tau=0\,,\\
\label{ec2l}
&&\partial_\tau\widetilde{T}^{0z}+
\partial_x\widetilde{T}^{zz}+2\widetilde{T}^{0z}/\tau=0\,,
\end{eqnarray}
where $dx=\tau\hsp d\eta$\,. These equations are equivalent to
Eqs.~(\ref{bcce1})--(\ref{emte2}) of Sect.~\ref{feqs}.  One can see
that Eqs.~(\ref{bccl})--(\ref{ec2l}) have the form (\ref{conf}) with
the replacement \mbox{$t\to\tau$}, \mbox{$z\to\tau\eta$},
\mbox{$v_z\to\rm{tanh}\hsp (Y-\eta)$}.

In the baryon--free case $\widetilde{J}^0=\widetilde{J}^z=0$ and
\re{bccl} is trivially satisfied. The ''phoenical'' version of SHASTA
algorithm~\cite{Bor73} is used to solve numerically
Eqs.~(\ref{ec1l})--(\ref{ec2l}) in the rectangle
\mbox{$|\eta|<\eta_{\rm max}$}, \mbox{$\tau_0<\tau<\tau_{\rm max}$}
of the $\eta-\tau$ plane.  Typically we choose \mbox{$\eta_{\rm max}=10$},
$\tau_{\rm max}=200$\,fm/c.  The grid size $\Delta\eta=1.25\cdot
10^{-2}$ is used in most calculations.  For each time step the
increment $\Delta\tau$ is found from the requirement
$\Delta\tau={\rm min}\hsp(0.1\tau\Delta\eta,\, 0.1\,{\rm fm/c})$\,.
The latter is motivated by the Courant condition for the stability of
fluid--dynamical equations in a finite difference approximation. Within
the light cone representation this conditions takes the form
$\Delta\tau<\Delta x/c_s=\tau\Delta\eta/c_s$\,, where $c_s<c$ is the
sound velocity.

\section{Particle spectra from resonance decay}

One can simplify the expressions (\ref{rdc}) for spectra of particles
produced in resonance decays, having in mind that degeneracy effects in
resonance phase--space densities should be small at small $\mu_F$ and
realistic temperatures $T_F\gtrsim 100$\,MeV.  Therefore, when
calculating invariant momentum distributions of resonances, one can
approximate the Bose or Fermi functions in \re{spec} by the
corresponding Boltzmann--J\"uttner expression:
\bel{rdf}
E_R\frac{d^{\hsp 3} N_R}{d^{\hsp 3} p_R}\simeq\frac{g_R}{(2\pi)^3}
\int d\sigma_\mu p_R^{\,\mu}\exp\left(\frac{\mu_F-p_R\hsp U_F}{T_F}\right)\,,
\ee
where $g_R$ is the degeneracy factor of the resonance $R$\,. Presence
of delta--function allows us to easily remove integration over the
angle between $\bm{p}_R$ and $\bm{p}$ in~\re{rdc}. Using further the
Lorentz--invariance of (\ref{rdc}), it is possible to make analytic
integration over the resonance energy in the fluid rest frame. The
result can be written as
\bel{rd1}
E\frac{d^{\hsp 3} N_{R\to i}}{d^{\hsp 3} p}=\frac{d^{\hsp 3} N_{R\to i}}
{d^{\hsp 2} p_T\hsp dy}=
\frac{g_Rb_{R\to i}}{16\pi^2}\int\limits_{\ds m_{\rm thr}}^\infty d\hsp m_R
\hsp\frac{m_R}{q_0}\hsp w\hsp (m_R)\int d\sigma_\mu
\left(A\hsp p^{\,\mu}+B\hsp U_F^\mu\right)\,.
\ee
The Lorentz--scalars $A$ and $B$ in the r.h.s. are determined as follows
\begin{eqnarray}
&&A=\frac{T_F}{p^{\,\prime 3}}\left[E^{\,\prime}
(E_-+T_F)-m_RE_{0\hsp i}\right]e^{\,\ds (\mu_F-E_-)/T_F}
-(E_-\to E_+)\,,\\
&&B=\frac{T_F}{p^{\,\prime 3}}\left[E^{\,\prime}
m_RE_{0\hsp i}-m_i^2(E_-+T_F)\right]e^{\,\ds (\mu_F-E_-)/T_F}
-(E_-\to E_+)\,,
\end{eqnarray}
where
$E^{\,\prime}=\sqrt{m_i^2+p^{\,\prime 2}}\equiv p\hsp U_F=m_T\cosh (y-Y_F)$
and $E_{\pm}\equiv m_R\hsp (E_{0\hsp i}E^{\,\prime} \pm q_0 p^{\,\prime})
/m_i^2$\,.

In most cases, when calculating contributions of resonance decays we
apply the zero--width approximation, i.e. substitute $w\hsp
(m_R)=\delta\hsp (m_R-\ov{m}_R)$\,, where $\ov{m}_R$ is the central
mass value\footnote
{
For resonances with several charged states we use the isospin--averaged
value.
}
presented in Ref.~\cite{PDG04} for the resonance $R$\,.  However, we
take into account the nonzero width of $\rho$--mesons. Following Ref.~\cite{Sol90},
we apply the parametrization
\bel{rmdi}
w\hsp (m_R)=\frac{C\gamma (m_R)}{(m_R^2-\ov{m}_R^{\,2})^2+\gamma^2(m_R)}\,,
\ee
Here the constant $C$ is taken from the normalization
condition, $\int\limits_{m_{\rm thr}}^\infty d\hsp m_R w\hsp (m_R)=1$\,, and
\bel{rmdi1}
\gamma(m_R)=\xi \frac{m_R\hsp q_0^3}{q_0^2+\Lambda^2}\,,
\ee
where $q_0=\sqrt{m_R^2/4-m_\pi^2}$\,.  The numerical
values~\cite{Sol90}
\bel{rmdi2}
\ov{m}_R=770\,{\rm MeV},~~\Lambda=159\,{\rm MeV},~~\xi=0.5\,,
\ee
are used to calculate feeding from $\rho$--meson decays.

\begin{acknowledgments}
The authors thank I.G. Bearden, J.J. Gaardh{\o}je, M.I. Gorenstein,
Yu.B.~Ivanov, L.~McLerran, D.Yu.~Pe\-ressounko, D.H. Rischke, and V.N.
Russkikh for useful discussions.  This work was supported in part by
the BMBF, GSI, the DFG grant~~436 RUS ~~~\mbox{113/711/0--2} (Germany)
and the grants RFBR 05--02--04013 and NS--8756.2006.2 (Russia).
\end{acknowledgments}

\end{document}